\documentclass[aps,prd,twocolumn,titlepage,superscriptaddress,nofootinbib]{revtex4}
\usepackage{graphicx}% Include figure files
\usepackage{color}
\usepackage{latexsym}
\usepackage{cancel}

\usepackage[normalem]{ulem}
\usepackage{hyperref,amssymb}
\usepackage{url}
\usepackage{color,soul}
\usepackage{graphicx}
\usepackage{verbatim}
\usepackage{multirow}
\usepackage{amsmath}
\newcommand{\beq}{\begin{eqnarray}}
	\newcommand{\eeq}{\end{eqnarray}}
\usepackage{mathrsfs}
\usepackage{float,soul}
\usepackage[dvipsnames]{xcolor}
\usepackage{mathtools}
\usepackage{slashed}
\usepackage{physics}	% installed packages for typing maths and physics
\usepackage{graphicx}   % need for Fig.ures
\usepackage{epstopdf}
\usepackage{subfigure}  % use for side-by-side Fig.ures
\usepackage{hyperref}   % use for hypertext links, including those to external documents and URLs
\usepackage{bbold}
\usepackage{wasysym}
\usepackage{feynmp}
\usepackage{hyperref}
\hypersetup{colorlinks,citecolor=blue}

\makeatletter
\newcommand{\myBig}{\bBigg@{1.75}}
\makeatother
\begin{document}

	\title{\Large Phase transitions in a holographic superfluid model with non-linear terms beyond the probe limit} 
	
	\author{Zi-Qiang Zhao}%\email{zhaoziqiang@stumail.neu.edu.cn}
	\affiliation{Key Laboratory of Cosmology and Astrophysics (Liaoning), College of Sciences, Northeastern University, Shenyang 110819, China}
	\author{Zhang-Yu Nie}\email{niezy@kust.edu.cn}
	\affiliation{Center for Gravitation and Astrophysics, Kunming University of Science and Technology, Kunming 650500, China}
	\author{Jing-Fei Zhang}%\email{jfzhang@mail.neu.edu.cn}
	\affiliation{Key Laboratory of Cosmology and Astrophysics (Liaoning), College of Sciences, Northeastern University, Shenyang 110819, China}
	\author{Xin Zhang}\email{zhangxin@mail.neu.edu.cn}
	\affiliation{Key Laboratory of Cosmology and Astrophysics (Liaoning), College of Sciences, Northeastern University, Shenyang 110819, China}
	\affiliation{Key Laboratory of Data Analytics and Optimization for Smart Industry (Ministry of Education), Northeastern University, Shenyang 110819, China}
	\affiliation{National Frontiers Science Center for Industrial Intelligence and Systems Optimization, Northeastern University, Shenyang 110819, China}

	\begin{abstract}
		We study the holographic s-wave superfluid model with fourth- and sixth-power self-interaction terms $\lambda |\psi|^4$ and $\tau |\psi|^6$, considering the full back-reaction of the matter fields on the metric in the 3+1 dimensional bulk. The self-interaction terms are effective at controlling the condensate to realize various phase transitions, such as zeroth-order, first-order, and second-order phase transitions within the single condensate s-wave superfluid model. Therefore, in this work, we investigate the influence of the back-reaction strength on various phase transitions, including zeroth-order and first-order phase transitions.   
		In addition, we confirm that the influence of the fourth- and sixth-power terms on the superfluid phase transition in the case of finite back-reaction is qualitatively the same as in the probe limit, thus presenting universality. We also plot the special values $\lambda_s$ of the parameter $\lambda$ at different back-reaction strengths, below which the condensate grows in the opposite direction. These values are important in controlling the order of the superfluid phase transitions. Comparing the influence of the back-reaction parameter with that of the higher-order non-linear coefficients, we see that the back-reaction strength brings in effective couplings similar to both the fourth-power and sixth-power terms.
		
	\end{abstract}
	\maketitle
	\section{Introduction}
	%From the holographic principle to the large $N$ limit, 
	The anti-de Sitter (AdS)/conformal filed theory (CFT) correspondence, introduced by Maldacena \cite{Maldacena:1997re} in 1997, reveals an intrinsic connection between field theory and gravitational theory and has provided a new perspective on understanding the nature of gravity and quantum matter. As a strong/weak duality, it is also regarded as a possible solution to strongly coupled problems.  
	One of the most important applications of the AdS/CFT correspondence is the description of superconductor phase transitions, also known as the holographic superconductor model or the Hartnoll-Herzog-Horowitz (HHH) model \cite{Hartnoll:2008vx,Hartnoll:2008kx,Herzog:2010vz}. The HHH model describes a simple holographic s-wave superconductor that undergoes a second-order phase transition below the critical temperature $T_c$. Further studies have shown that in holographic systems, one can also realize p-wave \cite{Gubser:2008wv,Cai:2013aca} and d-wave \cite{Chen:2010mk,Kim:2013oba} superconductors, as well as the coexistence and competition between multiple order parameters \cite{Basu:2010fa,Musso:2013ija,Nie:2013sda,Donos:2013woa,Li:2017wbi,Nie:2015zia,Nie:2014qma,Amado:2013lia,Zhang:2021vwp}. Additionally, various gravity backgrounds have been explored in holographic superconductor models \cite{Qiao:2020hkx,Pan:2021jii,Zhang:2023uuq,Ghorai:2021uby,Cai:2009hn}. As they offer complete duality that helps us understand strongly coupled systems, holographic models not only allow the study of equilibrium solutions but also provide insights into non-equilibrium dynamic evolution \cite{Zeng:2022hut,Xia:2023pom,Xia:2021jzh,Zhao:2023ffs,Su:2023vqa}. The AdS/CFT correspondence has also been used to study novel phenomena such as supersolids \cite{Baggioli:2022aft,Baggioli:2022pyb,Baggioli:2019rrs,Baggioli:2021tzr,Baggioli:2022uqb} and amorphous solids \cite{Pan:2021cux}.
	
	Another application of the AdS/CFT correspondence is to help us better understand the nature of black holes and the underlying laws of quantum gravity. As one of the most mysterious yet simplest objects in the universe, black holes have always been critical to the study of gravity. Despite extensive studies on black holes, some of their physical phenomena remain beyond the current theoretical framework, such as black hole singularities, jets, and the black hole information paradox~\cite{Witten:2024upt}. Additionally, black holes can serve as standard probes in cosmology to investigate our universe. For instance, gravitational waves produced by black hole mergers are often referred to as dark sirens \cite{Chen:2017rfc,DES:2019ccw,Zhao:2019gyk,LIGOScientific:2021aug,Song:2022siz,Jin:2023sfc,Jin:2023tou,Dong:2024bvw,Xiao:2024nmi,Song:2025ddm}. It is very important to study the dynamics of the metric fields' gravitational systems. Therefore, it is urgent and meaningful to extend the studies of holographic superfluids from the probe limit, where the metric is fixed as the background, to the complete case including the full back-reaction of matter fields on the metric~\cite{Hartnoll:2008kx,Cai:2013aca,Pan:2021jii,Cai:2013wma,Wang:2016jov,Wang:2019vaq}. 
	The interplay between the matter and metric fields renders the phase transition behavior of holographic superfluids richer and more complex, introducing phenomena such as reentrant phase transitions~\cite{Nie:2014qma}, first-order phase transitions~\cite{Ammon:2009xh,Cai:2013aca,Cai:2014ija,Nie:2014qma}, and zeroth-order phase transitions~\cite{Cai:2013aca,Cai:2014ija}.
	
	%此外，我们还可以用全息对偶来研究相变。
	In a recent study on the holographic s-wave superfluid model~\cite{Zhao:2022jvs}, the authors used fourth- and sixth-power terms to achieve powerful control over the superfluid phase transitions in the probe limit, where usually only second-order phase transitions are realized. They explored the parameter space of this model, where various phase transitions appear, including second-order, first-order, zeroth-order, as well as ``cave of wind'' (COW) phase transitions. The linear stability of these phase transitions was also studied by calculating the quasinormal modes (QNMs), which confirmed the landscape deduced from the free energy of the on-shell states. Later, the full dynamical spinodal decomposition process in the same model was realized in Ref.~\cite{Zhao:2023ffs}, which confirmed the inhomogeneous linear instability from the QNMs and observed the non-equilibrium creation and evolution of bubbles. This same simple model has also been utilized to study problems in the supercritical region~\cite{Zhao:2024jhs}.

	%However, the role of the back-reaction in such systems remains unclear. 
	From the above-described progress, it is interesting to study the simple s-wave superfluid model with fourth- and sixth-power terms while considering the back-reaction on the metric. On one hand, we can explore how the back-reaction strength controls various phase transitions, and on the other hand, we can test the universal control of the non-linear terms on phase transitions. Finally, this setup provides a convenient way to compare the influence of the back-reaction parameter with that of the non-linear terms, thereby promoting better understanding of the effect of metric dynamics in holographic phase transitions.

	In this paper, we investigate the holographic superfluid model containing self-interaction terms while considering the full dynamics involving both metric and matter fields. 
	The rest of this paper is structured as follows. In Sect. \ref{sec2}, we present the holographic setup and the details of the calculations. 
	In Sect. \ref{sec3}, we study how the back-reaction strength controls various phase transitions. In Sect. \ref{sec4}, we examine the universal control of the fourth- and sixth-power terms on phase transitions at finite back-reaction strength and provide the dependence of the special value $\lambda_s$ on the back-reaction strength. 
	Finally, we present some conclusions and discussions in Sect. \ref{secConclusions}.

	\section{The holographic model}\label{sec2}
	%我们在之前的工作中研究了含有两个非线性项的全息超流模型，通过改变自耦和项的参数lambda和tau，我们可以得到zeroth，first和second相变。在这个工作中，我们研究了物质场反作用对于s波全息超导模型的影响。我们从作用量出发.
	
	%介词后面要加动名词
	\subsection{Equations of motion}
	In Ref.~\cite{Zhao:2022jvs}, the second-order, first-order, and zeroth-order phase transitions are easily realized by considering the fourth- and sixth-power scalar potential terms in the probe limit. In order to investigate the back-reaction of matter fields on the metric, we consider the following action including both the matter part $S_M$ and the gravity part $S_G$
	\begin{align}
		&S=\,S_{M}+S_{G}~,\quad
		S_G=\,\frac{1}{2\kappa_g ^2}\int d^{4}x\sqrt{-g}\left(R-2\Lambda\right)~,\label{Lagg}\\
		&S_M=\,\frac{1}{q ^2}\int d^{4}x\sqrt{-g}(-\frac{1}{4}F_{\mu\nu}F^{\mu\nu}
		-D_{\mu}\psi^{\ast}D^{\mu}\psi  \nonumber\\
		&~~~~~~~~~~-m^{2}\psi^{\ast}\psi-\lambda(\psi^{\ast}\psi)^{2}-\tau(\psi^{\ast}\psi)^{3})~.\label{Lagm}
	\end{align}
	Here, $F_{\mu\nu}=\nabla_{\mu}A_{\nu}-\nabla_{\nu}A_{\mu}$ is the Maxwell field strength and $D_{\mu}\psi=\nabla_{\mu}\psi-i A_\mu\psi$ is the standard covariant derivative term of the charged scalar filed $\psi$. 
	
	The Einstein equation is
	\begin{align}
		R_{\mu\nu}-\frac{1}{2}(R-2\Lambda)g_{\mu\nu}=b^2\mathcal{T}_{\mu\nu},
	\end{align}
	where $b=\kappa_g/q$ describes the strength of back-reaction of matter fields on the background geometry and $\kappa_g^2=8\pi G$. $\mathcal{T}_{\mu\nu}$ is the stress-energy tensor of the matter fields
	\begin{align}
		\mathcal{T}_{\mu\nu}=&(-\frac{1}{4}F_{\alpha\beta}F^{\alpha\beta}
		-D_{\alpha}\psi^{\ast}D^{\alpha}\psi-m^{2}\psi^{\ast}\psi\nonumber\\
		&-\lambda(\psi^{\ast}\psi)^{2}-\tau(\psi^{\ast}\psi)^{3})g_{\mu\nu}+(D_{\mu}\psi^{\ast}D_{\nu}\psi\nonumber\\
		&+D_{\nu}\psi^{\ast}D_{\mu}\psi)+F_{\mu\alpha}F^{\alpha}_{\nu}.
	\end{align}
	
	We use the standard ansatz for realizing the holographic superfluid phase transition
	\begin{align}
		\psi=\psi(r)~,~A_\mu dx^\mu=\phi(r)dt~,
	\end{align}
	and the line element is consistently set to
	\begin{align}
		ds^{2}=-N(r)\sigma(r)^2dt^{2}+\frac{1}{N(r)}dr^{2}+r^{2}dx^{2}+r^{2}dy^{2},
	\end{align}
	with
	\begin{align}
		N(r)=\frac{r^2}{L^2}(1-\frac{2M(r)}{r^3}),
	\end{align}
	where $L$ is the AdS radius. The Hawking temperature of such a black brane spacetime is
	\begin{align}
		T=\frac{N'(r_h)\sigma(r_h)}{4 \pi},
	\end{align}
	where $r=r_h$ labels the position of the event horizon.
	
	With the above ansatz, the full equations of motion can be written as
	\begin{align}
		\psi''(r)=&\left(\frac{m^{2}}{N(r)}-\frac{\phi(r)^{2}}{N(r)^2\sigma(r)^2}\right)\psi(r)+\frac{2\lambda}{N(r)}\psi(r)^{3}\nonumber\\
		&+\frac{3\tau}{N(r)}\psi(r)^{5}-\left(\frac{N'(r)}{N(r)}+\frac{2}{r}+\frac{\sigma'(r)}{\sigma(r)}\right)\psi'(r),\label{eqpsi}\\
		\phi''(r)=&\left(\frac{\sigma'(r)}{\sigma(r)}-\frac{2}{r}\right)\phi'(r)+\frac{2\psi(r)^{2}}{N(r)}\phi(r),\label{eqphi}\\
		\sigma'(r)=&b^2r\left(\frac{\phi(r)^2\psi(r)^2}{N(r)^2\sigma(r)}+\sigma(r)\psi'(r)^2\right),\label{eqsigma}\\
		M'(r)=&\frac{b^2r^2}{\sigma(r)^2}\left( \frac{L^2\phi(r)^2\psi(r)^2}{2~N(r)}+\frac{L^2\phi'(r)^2}{4} \right)\nonumber\\
		&+\frac{1}{2}b^2r^2L^2\left(N(r)\psi'(r)^2+m^2\psi(r)^2\right)\nonumber\\
		&+\frac{1}{2}L^2b^2r^2\left(\lambda\psi(r)^4+\tau\psi(r)^6\right)~,\label{eqn}
	\end{align}
	with three sets of scaling symmetries:
	\begin{align}
		&1.~\phi\rightarrow a^{2}\phi~,~\psi \rightarrow a\psi~,~N\rightarrow a^2N~,~m\rightarrow a m~,\nonumber\\
		&\quad\quad L\rightarrow a^{-1}L~,
		b\rightarrow a^{-1}b~,~\tau\rightarrow a^{-2}\tau.\nonumber\\
		&2.~\phi\rightarrow a\phi~,~\sigma \rightarrow a\sigma.\nonumber\\
		&3.~\phi\rightarrow a\phi~,~N\rightarrow a^2 N,~M\rightarrow a^3 M~,~r\rightarrow ar.\nonumber\\
		%&(4).~\phi\rightarrow a\phi~,~\psi\rightarrow a\psi~,~b\rightarrow a^{-1}b~,~q\rightarrow a^{-1}q~,~\lambda\rightarrow a^{-2}\lambda~,\nonumber\\
		%&\quad\quad\tau\rightarrow a^{-4}\tau.
		\label{scaling}
	\end{align}
	
	%我们将用数值的方法求解上述所有方程。为了求解微分方程，我们需要考虑视界和边界处场的行为。
	%上述方程中，N和σ都是一阶微分方程，所以都只需要一个边界条件，psi和phi则和探子极限下的s波超导模型相同。
	%在视界处，方程的展开形式为。因为上述scaling symmetry的存在，
	In order to solve these equations, we need to specify boundary conditions on both the horizon $r=r_h$ and the asymptotic boundary as $r\rightarrow\infty$. Without loss of generality, we set $L=1$ and $G=1$ for the rest of this paper. We also set $r_h=1$ in numerical calculations and recover the value of $r_h$ using the scaling symmetry $3$, while scaling either the chemical potential $\mu$ or the charge density $\rho$ to a fixed value. The expansions of the functions near the horizon are
	%The equations for $M(r)$ and $\sigma(r)$ are first order while the equations for $\phi$ and $\psi$ are second order, therefore we need totally 6 boundary conditions to fix a single solution.
	\begin{align}
		\phi(r)=&\phi_{h_1}(r-r_h)+\phi_{h_2}(r-r_h)^2+\cdots~,\\
		\psi(r)=&\psi_{h_0}+\psi_{h_1}(r-r_h)+\cdots~,\\
		\sigma(r)=&\sigma_{h_0}+\sigma_{h_1}(r-r_h)+\cdots~,\\
		M(r)=&\frac{r^3_h}{2}+M_{h_1}(r-r_h)+\cdots~.
	\end{align}
	The horizon condition requires $N(r_h)=0$, which is equal to $M(r_h)=r^3_h/2$. Near the AdS boundary, the expansions of the functions are
	\begin{align}
		\phi(r)&=\mu-\frac{\rho}{r}+\cdots~,\\
		\psi(r)&=\frac{\psi^{(1)}}{r}+\frac{\psi^{(2)}}{r^2}+\cdots~,\\
		\sigma(r)&=\sigma_{b_0}+\frac{\sigma_{b_3}}{r^3}+\cdots~,\\
		M(r)&=M_{b_0}+\frac{M_{b_1}}{r}+\cdots~.
	\end{align}
	
	The scaling symmetry (\ref{scaling}) will be used to rescale any solution to be asymptotically AdS, which means %$\mathop{lim}\limits_{r\rightarrow\infty}\sigma(r)\rightarrow 1$
	$\sigma(\infty)=1$. The remaining boundary conditions are: $\psi(r_h)=-(3/2)\psi'(r_h)+M'(r_h)\psi'(r_h)$, $\psi^{(1)}(\infty)=0$, $\phi(r_h)=0$, $\phi(\infty)=\mu$, where $\mu$ is the chemical potential. With these boundary conditions, we can solve the equations of motion numerically. 
	
	%在这篇文章中，我们选择和（）一样的标准量子化方案。因为我们的模型只含有s波序参量，所以<o>等于psi（2）。反作用会同时影响临界温度和临界mu，所以我们选择t为自由参数，并且固定mu。后续章节中，我们绘图用的都是无量纲参数t/mu。
	In this paper, we choose the standard quantization scheme, which means $\psi^{(1)}=0$ and the non-trivial vacuum expectation value is $\langle \mathcal{O} \rangle=\sqrt{2}\psi^{(2)}$. 
	
	\subsection{Free energy}
	In this work, we fix the chemical potential $\mu$ and obtain solutions at various temperatures $T$, which means we are working in the grand canonical ensemble.
	An essential tool for confirming the order of a phase transition in the grand canonical ensemble is the grand potential. In this section, therefore, we  provide the formula for calculating the grand potential density. The grand potential of this holographic system equals the temperature times the Euclidean on-shell action in the bulk spacetime \cite{Cai:2013aca,Nie:2014qma}
	\begin{align}
		\Omega=TS_E~,
	\end{align}
	where $S_E$ is
	\begin{align}
		S_E=&\frac{1}{2\kappa_g^2}\int d^{4}x\sqrt{-g}\left(R-2\Lambda+2b^2\mathcal{L}_{matter}\right)\nonumber\\
		&-\frac{1}{\kappa_g^2}\int d^{3}x\sqrt{-h}\left(K+\frac{2}{L}  \right)~.\label{actionSe}
	\end{align}
	In Eq.~(\ref{actionSe}), $\mathcal{L}_{matter}$ is the Lagrangian density of the matter part and $K$ is the trace of the extrinsic curvature $K_{\mu\nu}$ for the boundary (see, \textit{e.g.}, Refs.~\cite{Cai:2002mr,Nie:2014qma}), where $K_{\mu\nu}=-h_\mu^{~\rho}\nabla_\rho n_\nu$ and $n$ is the unit normal to the boundary surface. By substituting the equations of motion into the action, one can derive the formula for the grand potential density at the boundary,
	\begin{align}
		\frac{2\kappa_g^2\Omega}{V_2}=&\mathop{lim}\limits_{r\rightarrow\infty}[2rN(r)\sigma(r)+r^2\sigma(r)N'(r)\nonumber\\
		&+2r^2N(r)\sigma'(r)-4r^2\sqrt{N(r)}\sigma(r)]~,
	\end{align}
	where $V_2$ is the area of the two dimensional transverse space. For the normal phase, the temperature and grand potential density are
	\begin{align}
		T=\frac{r_h}{4\pi}(3-\frac{b^2\mu^2}{2r_h^2})~,~\frac{2\kappa_g^2\Omega}{V_2}=-r_h^2-\frac{1}{2}b^2\mu^2r_h,
	\end{align}
	while for the condensed phases, they are
	\begin{align}
		&T=\frac{r_h}{4\pi}(3\sigma_{h_0}-\frac{b^2\phi_{h_1}^2}{2\sigma_{h_0}}-b^2m^2 \sigma_{h_0}\psi_{h_0}^2\nonumber\\ 
		&\quad~~~-b^2~\lambda~\sigma_{h_0}\psi_{h_0}^4-b^2~\tau~\sigma_{h_0}\psi_{h_0}^6)~,\nonumber\\
		%\frac{2\kappa_g^2\Omega}{V_2}=&-2M_{b_0}~.   
		&\frac{2\kappa_g^2\Omega}{V_2}=-2M_{b_0}~.
	\end{align}

	When $b=0$, the spacetime background decouples from the matter fields, leading us back to the probe limit where various phase transitions have been realized in previous studies (see,  \textit{e.g.}, Refs. \cite{Zhao:2022jvs,Zhao:2024jhs}).

	\section{The power of back-reaction strength on various phase transitions}\label{sec3}

	\subsection{Second-order phase transition}
	
	\begin{figure}[t]
		\center
		\includegraphics[width=0.8\columnwidth]{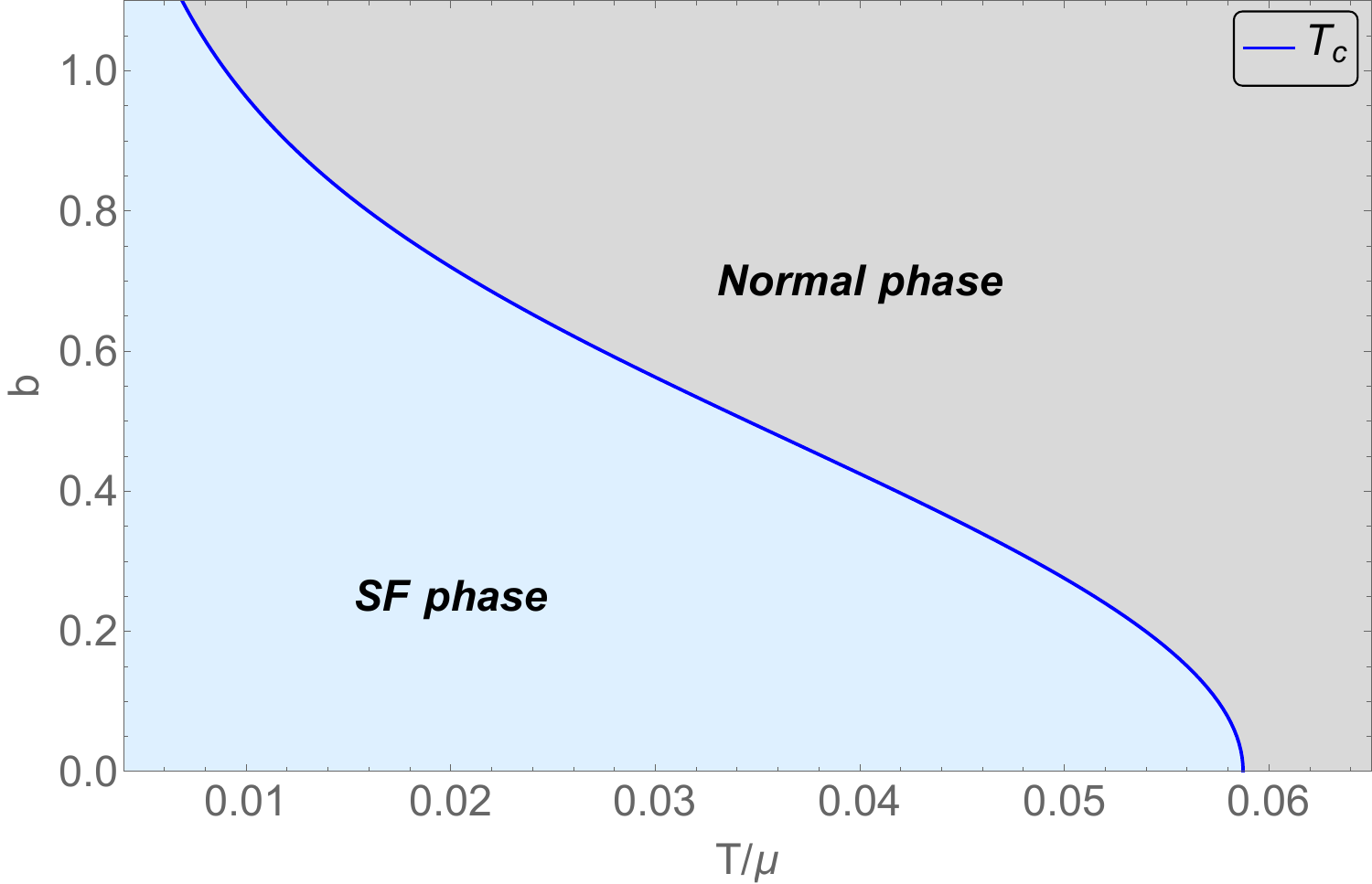}
		\caption{The $b-T$ phase diagram of temperature with $\lambda=0$ and $\tau=0$. The blue solid line represents the critical points of the second-order superfluid phase transitions.}\label{tcb}
	\end{figure}
	In this model, when the coefficients of the self-interaction terms $\lambda$ and $\tau$ are both equal to zero, only second-order phase transitions are realized. Since the back-reaction effect in this model was previously studied in Ref.~\cite{Hartnoll:2008kx}, we present only the $b-T$ phase diagram in Fig.~\ref{tcb}, which shows the relationship between the critical temperature $T_c$ and the back-reaction parameter $b$. It shows that the critical temperature $T_c$ decreases with the increase of the back-reaction parameter $b$. This feature tells us that the larger the back-reaction of matter fields on the background geometry, the more difficult it will be for the superfluid condensate to form.
	
	As the back-reaction increases, the critical temperature gradually approaches zero~\cite{Hartnoll:2008kx}, and simultaneously, the calculation of the condensate curve becomes more difficult. This numerical difficulty limits us to considering only a maximum value of $b=1.2$ for the back-reaction parameter. It is worth noting that the change in the back-reaction parameter not only affects the critical temperature but also alters the condensate curves.
	
	\subsection{Zeroth-order phase transitions and the power of $\tau$}\label{subsection0th}
\begin{figure}[t]
		\center
		\includegraphics[width=0.475\columnwidth]{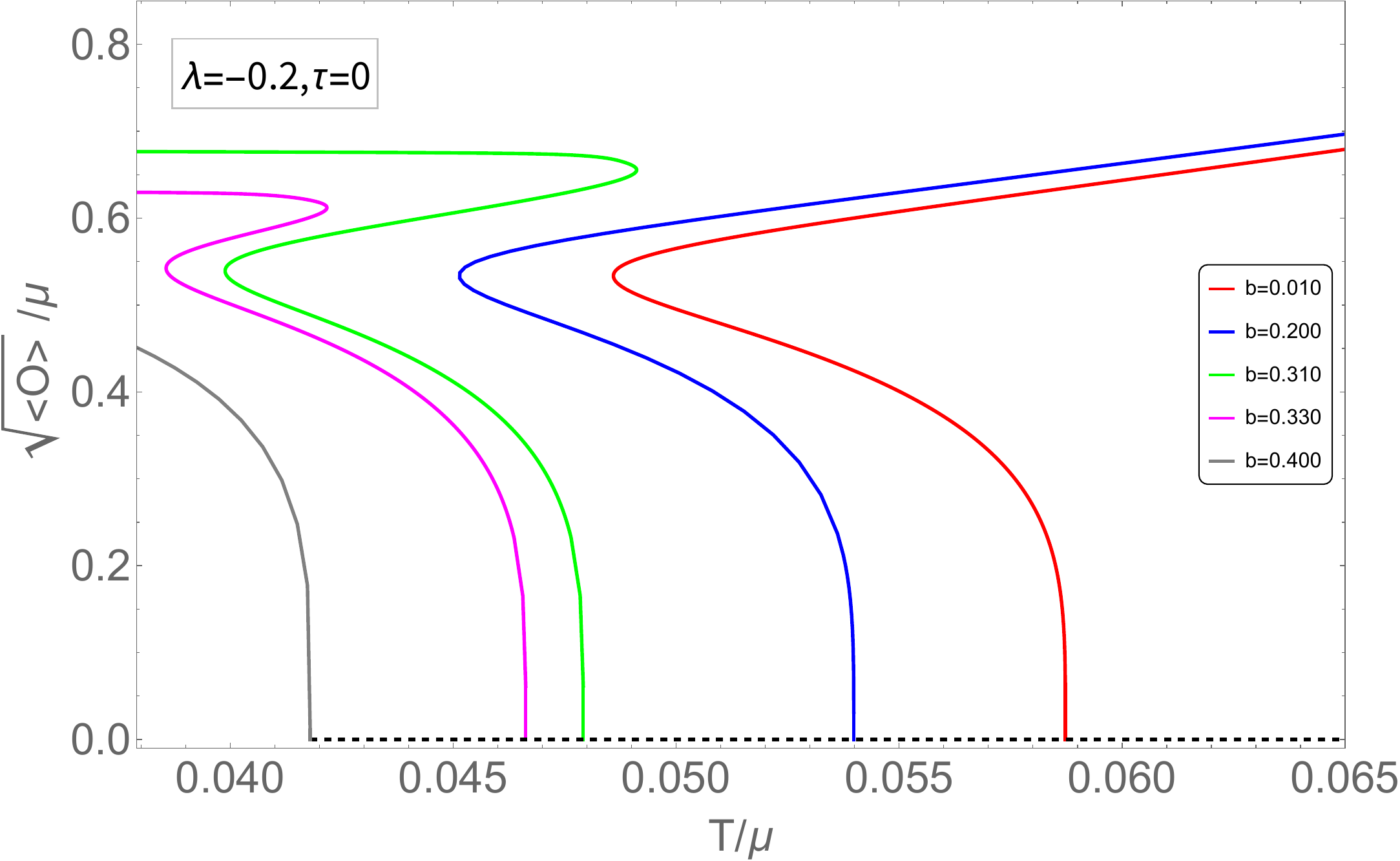}
		\includegraphics[width=0.49\columnwidth]{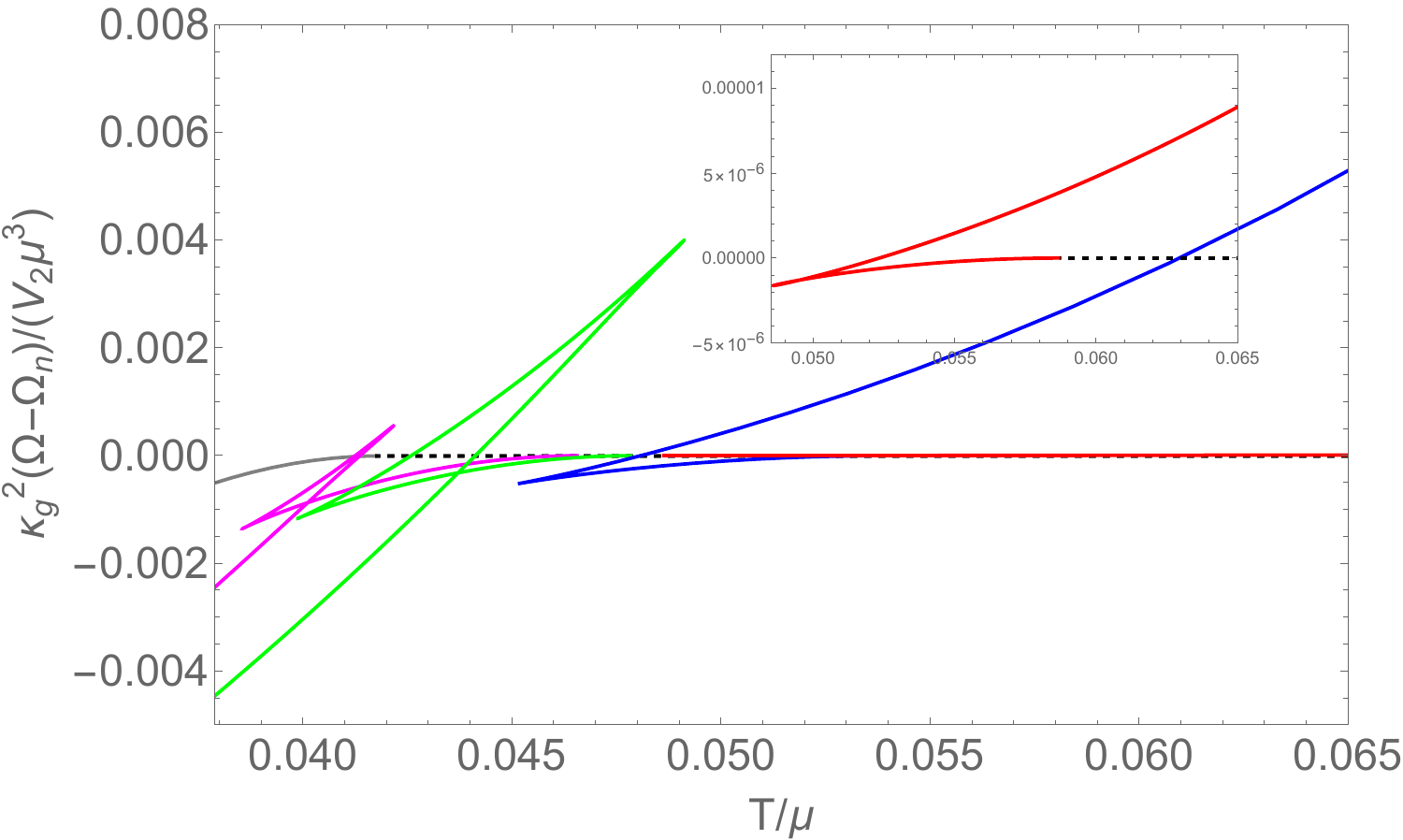}
		\caption{The dependence of the condensates as well as grand potential curves on the back-reaction strength $b$ with $\lambda=-0.2$ and $\tau=0$ . The left panel depicts the condensates, while the right panel represents the corresponding grand potential curves.
			The red, blue, green, magenta, and gray lines represent solutions with $b=0.010$, $0.200$, $0.310$, $0.330$, and $0.400$, respectively.}\label{lambda-02tau0}
	\end{figure}

	In the probe limit, with fixed parameters $\lambda = -0.2$ and $\tau = 0$, the system shows a zeroth-order superfluid phase transition~\cite{Zhao:2022jvs}. 
	We further introduce the back-reaction with $\lambda=-0.2$ to see how the zeroth-order phase transition changes with the increasing strength of the back-reaction. In Fig.~\ref{lambda-02tau0}, we illustrate the condensate curves with different values of the back-reaction parameter $b$ for $\lambda=-0.2$ and $\tau=0$. We see that as the back-reaction parameter $b$ increases, the phase transition gradually changes from a zeroth-order phase transition to a first-order phase transition and finally becomes a second-order phase transition.
	
	In Ref.~\cite{Zhao:2022jvs}, it has been discussed that when the self-interaction parameter $\lambda$ is less than zero, it implies an intrinsic attractive interaction for large values of the order parameter $\psi$, leading the system to suffer runaway instability while it undergoes a zeroth-order phase transition. This kind of instability in the zeroth-order phase transitions is confirmed from both the thermodynamic and dynamic perspectives~\cite{Zhao:2022jvs}. One simple method to rescue the system from this runaway instability is the introduction of an additional sixth-power self-interaction term with a positive value of $\tau$. This approach changes the system from a zeroth-order phase transition to a first-order phase transition with a small value of $\tau$ or a second-order phase transition with a very large value of $\tau$. Now we see a similar effect from the increasing value of back-reaction strength $b$.
	
	The green and magenta curves clearly contain the character of a first-order phase transition and the gray curve shows a typical second-order phase transition. Although the red and blue curves seem to be zeroth-order phase transitions, it is not confirmed whether the curves will turn back at large condensate values and become first-order phase transitions. 
	Therefore, we resort to the landscape analysis from the effective interaction of the back-reaction to confirm whether these phase transitions will turn into first-order phase transitions. 
	
	As discussed in Ref.~\cite{Zhao:2022jvs}, from the landscape perspective, it is clear that in the probe limit of the zeroth-order phase transition, as long as there is a positive $\tau$, no matter how small it is, it will always stabilize the system by bounding the thermodynamic potential landscape from below. %for sufficiently large values of $\langle \mathcal{O} \rangle$. 
	If the back-reaction on the metric brings in similar effective self-interaction, a condensate solution with very large values of $\langle \mathcal{O} \rangle$ always exists in the low temperature region, and the condensate curve will always turn back to form the standard first-order phase transition. Due to numerical limitations, we cannot obtain solutions with arbitrarily large values of $\langle \mathcal{O} \rangle$  to demonstrate this directly. However, we are able to confirm a similar such effective self-interaction of the back-reaction from the green, magenta, and gray curves in Fig.~\ref{lambda-02tau0}. Therefore, we suppose that the phase transitions with finite back-reaction strength, such as the red and blue cases in Fig.~\ref{lambda-02tau0}, indicate first-order phase transitions.
	%We believe that the results obtained through theoretical analysis are correct.
	Nevertheless, the back-reaction of the matter fields on the spacetime background exhibits more complex influences than the higher power non-linear potential terms. When the back-reaction parameter is changed, the critical temperature $T_c$ of the system changes simultaneously. In contrast, higher-order non-linear terms influence only the configuration of the condensate curves of the system without altering the critical temperature. 
	
	%但是我们相信，从理论上分析得到的结果是正确的。
	%尽管如此，物质场对时空背景的反作用仍然有不同的地方。当改变反作用参数时，系统的临界温度也会同时改变，但高阶非线性项只会改变系统的相变行为。
	
	\subsection{First-order phase transitions and the power of $\lambda$}\label{1storderlambda}
	%上面描述的情况中，全息模型在探子极限下不是二阶相变就是0阶相变，于是我们非常自然地就会想到，如果探子极限下系统就处于一阶相变呢？我们选取了一个探子极限处于一阶相变的参数，并逐渐增加反作用参数b，结果我们发现，这种情况下的结果与之前描述的结果并没有本质上的不同，并且我们在图5中展示了这种结果。我们为了能够更加清晰地展示结果，所以温度都除了Tc。正如我们前面讨论过的，一阶相变会随着反作用参数的增大，逐渐变成二阶相变。
	
	%事实上，我们还发现了另外一种与众不同的情况。当我们选取一个使得系统在b=0时就为二阶相变的非零的lambda和tau参数时，系统会先变成一阶相变，然后再回到二阶相变。值得注意的是，要得到这种奇特的变化趋势，我们需要选择一个非常接近临界lambdac的值。在b=0时，临界lambdac=0.757，为了保证系统没有不稳定的超流解，所以我们选择小于临界lambdac的参数，即lambda=。我们在图6中给出了这种情形下的凝聚图和自由能随着反作用的变化情况，并且在图7中给出了这种情况的相图，其中quasi-tc指的是没有自耦合项时候的临界温度Tc。
	In Sect.~\ref{subsection0th}, we confirmed that the back-reaction brings in effective interactions similar to the sixth (or higher) power potential term. To confirm whether the back-reaction also brings in effective interactions similar to the fourth-power term, which is important in switching the superfluid phase transitions between the second-order and first-order phase transitions~\cite{Zhao:2022jvs}, we set the value of $\lambda$ to be close to the special value $\lambda_s = -0.757$ in the probe limit, which controls whether the condensate curve grows leftwards or rightwards at the critical point. Then we see whether the growth direction changes with the back-reaction strength.
	\begin{figure}[t]
		\center
		\includegraphics[width=0.46\columnwidth]{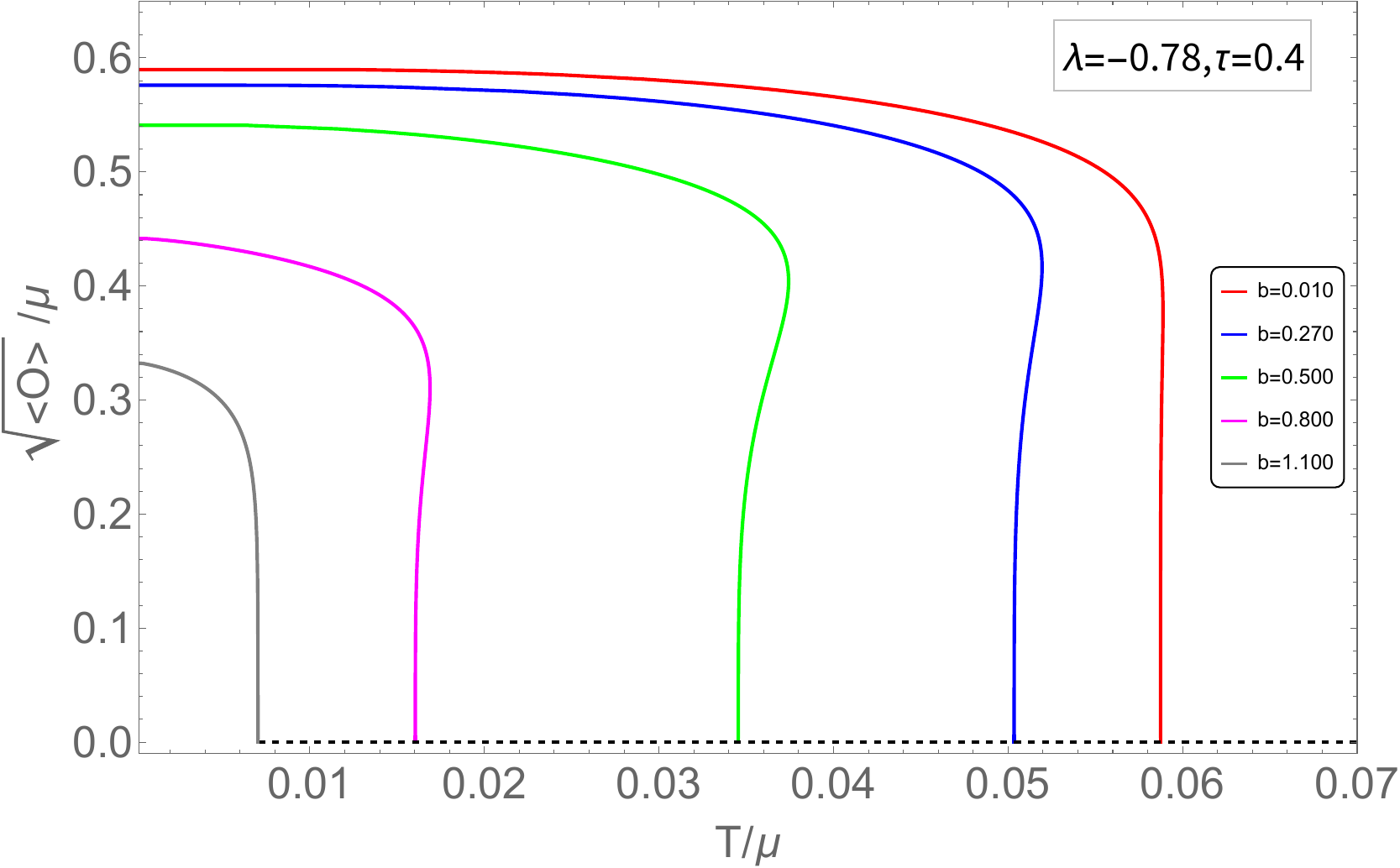}
		\includegraphics[width=0.49\columnwidth]{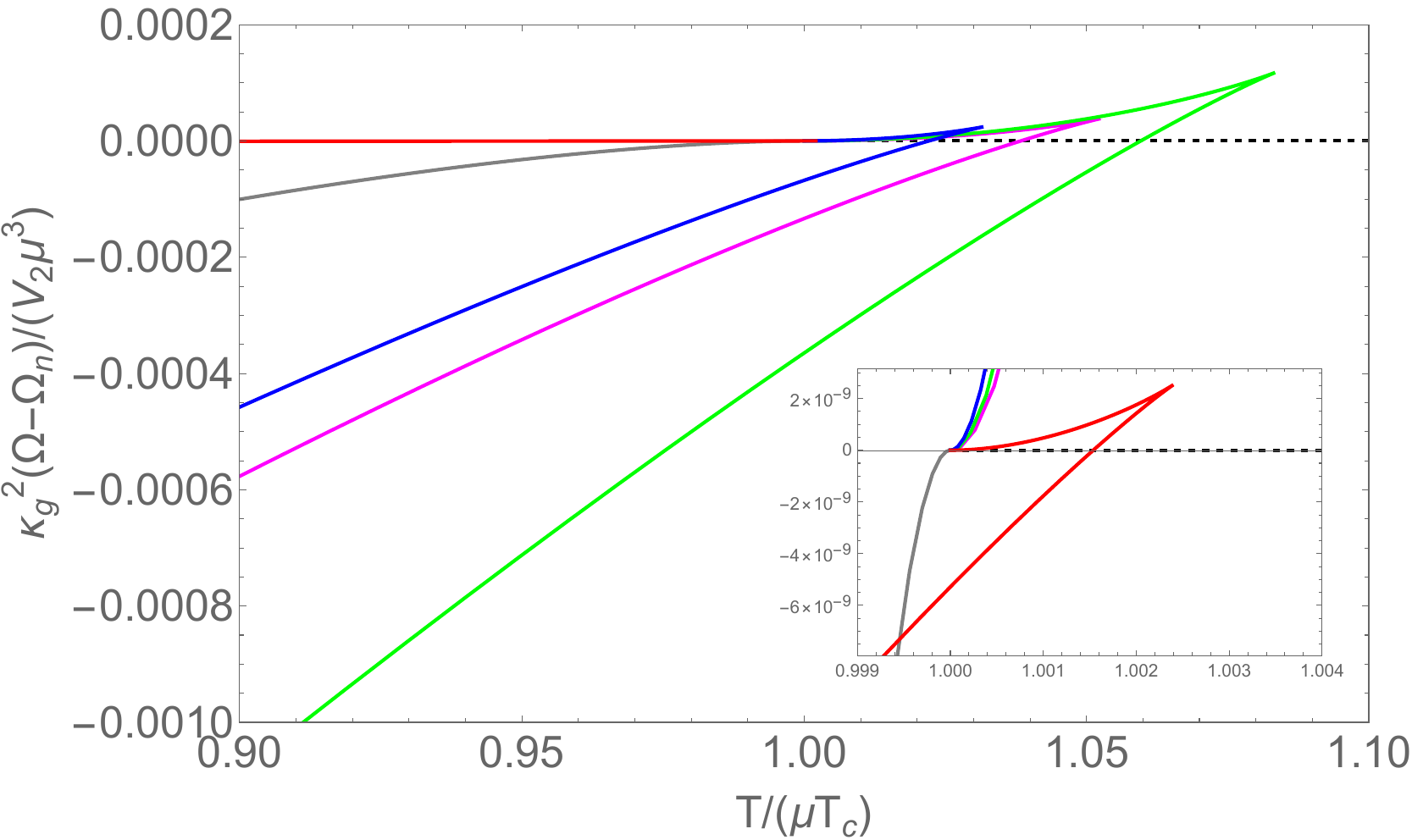}
		\caption{The dependence of the condensates as well as grand potential curves on the back-reaction strength $b$ with parameters $\lambda=-0.78$ and $\tau=0.4$. The left panel depicts the condensates, while the right panel represents the corresponding grand potential curves. The red, blue, green, magenta, and gray lines represent solutions with $b=0.010$, $0.270$, $0.500$, $0.800$, and $1.100$, respectively.}\label{lambda-0.78b0.4}
	\end{figure}
	
	We first set $\lambda=-0.78$ and $\tau=0.4$, where $\lambda$ is slightly lower than the special value $\lambda_s$ in the probe limit. In this case, the phase transition in the probe limit is first-order. We further present the condensates as well as the grand potential curves with increasing values of the back-reaction strength in Fig.~\ref{lambda-0.78b0.4}.
	It is clear that along with the increasing value of the back-reaction strength, the superfluid phase transition remains first-order with the decrease in the quasi-critical point.  To show the first-order phase transitions more clearly, we normalize the temperature with respect to $T_c$ in the right panel for the grand potential curves.
	
	%It is clear that along with the increasing value of the back-reaction strength, the superfluid phase transition remains first-order with the decreasing of the quasi-critical point.  To show the first-order phase transitions more clearly, we normalized the temperature with respect to $T_c$ in the right panel for the grand potential curves.
	
\begin{figure}[t]
		\center
		\includegraphics[width=0.46\columnwidth]{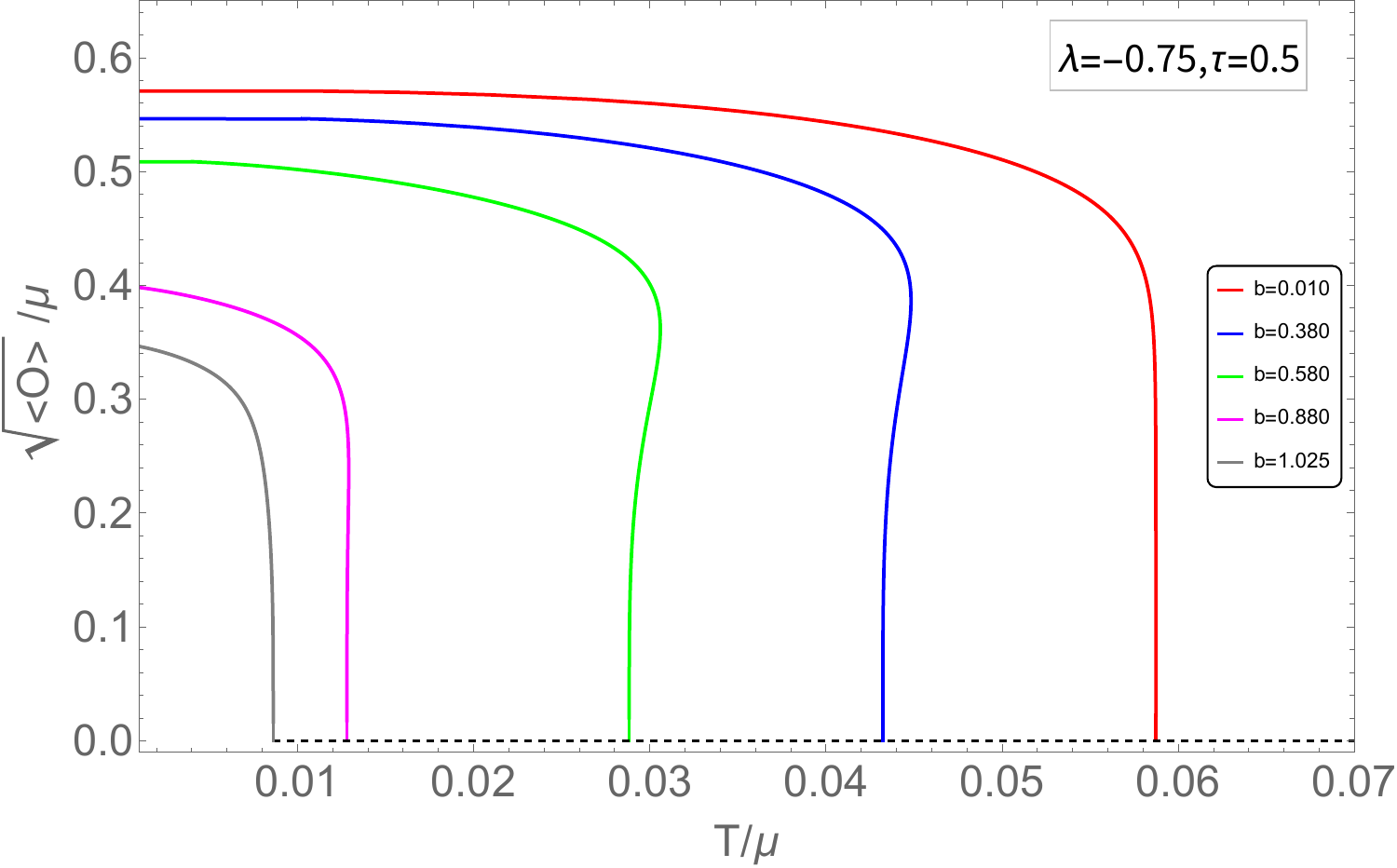}
		\includegraphics[width=0.49\columnwidth]{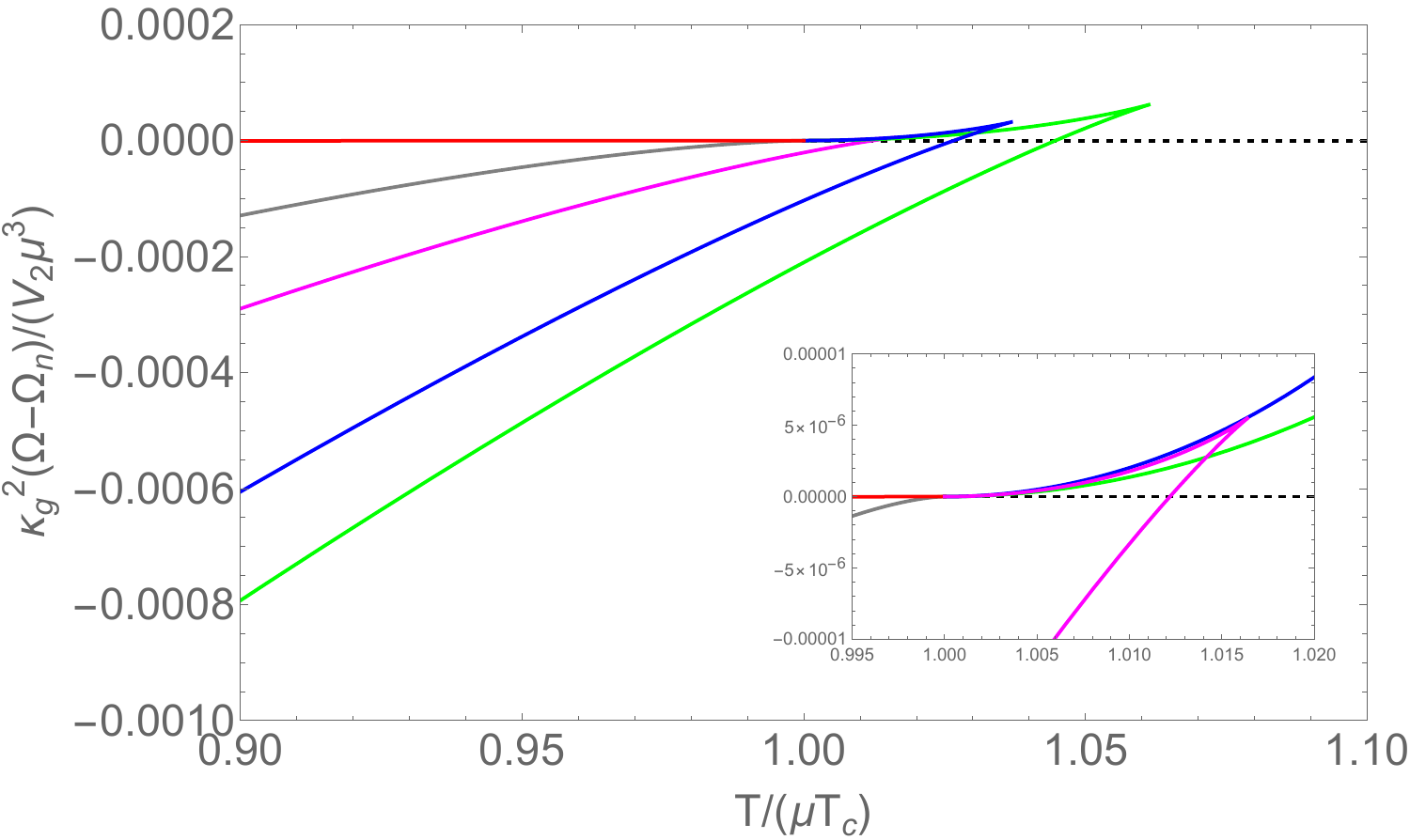}
		\caption{The dependence of the condensates as well as grand potential curves on the back-reaction strength $b$ with $\lambda=-0.75$ and $\tau=0.5$. The left panel depicts the condensates, while the right panel represents the corresponding grand potential curves. The red, blue, green, magenta, and gray lines represent solutions with $b=0.010$, $0.380$, $0.580$, $0.880$, and $1.025$, respectively. }\label{lambda-0.75b0.5}
	\end{figure}

\begin{figure}[t]
		\center
		\includegraphics[width=0.45\columnwidth]{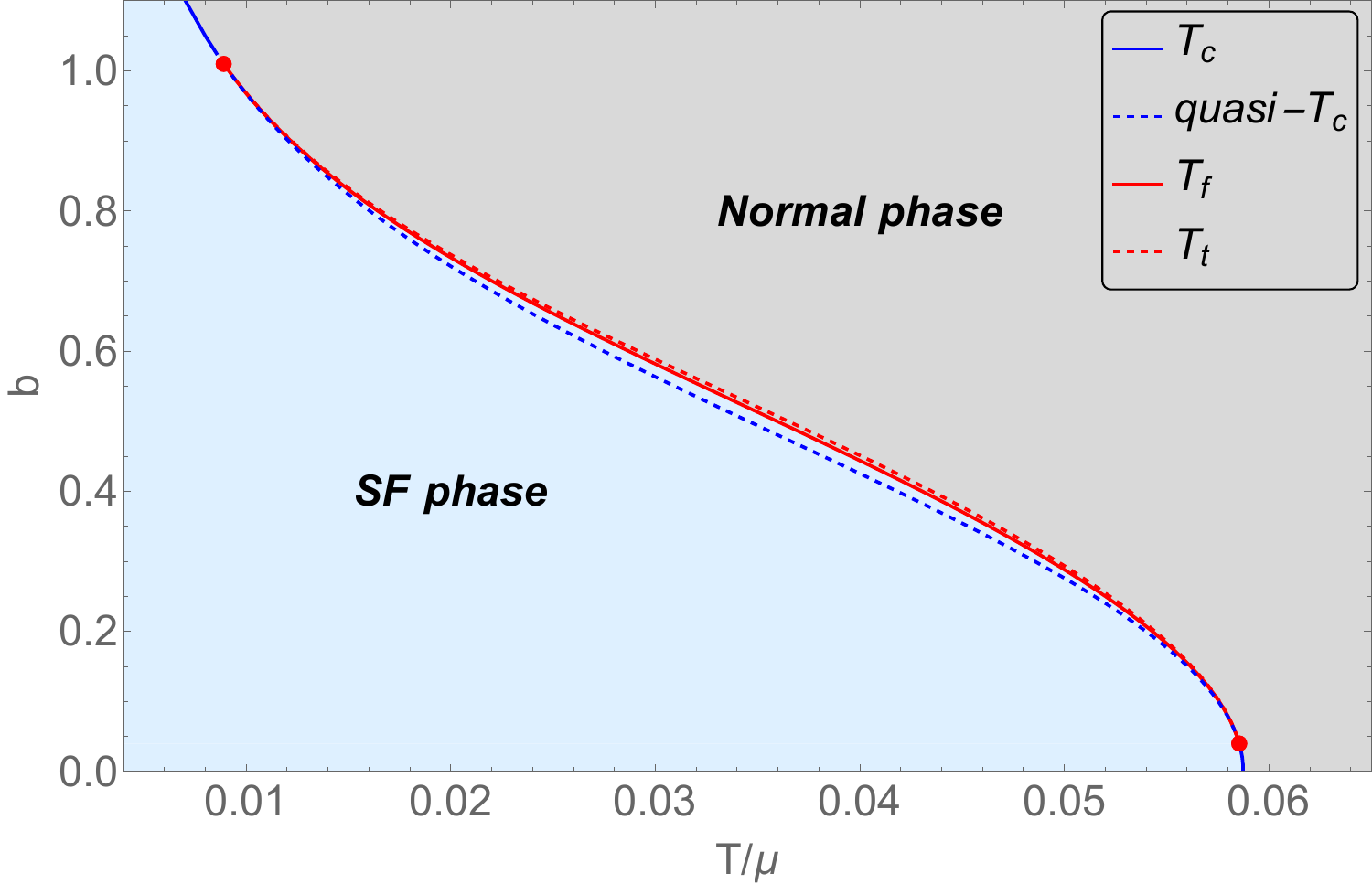}
		\includegraphics[width=0.45\columnwidth]{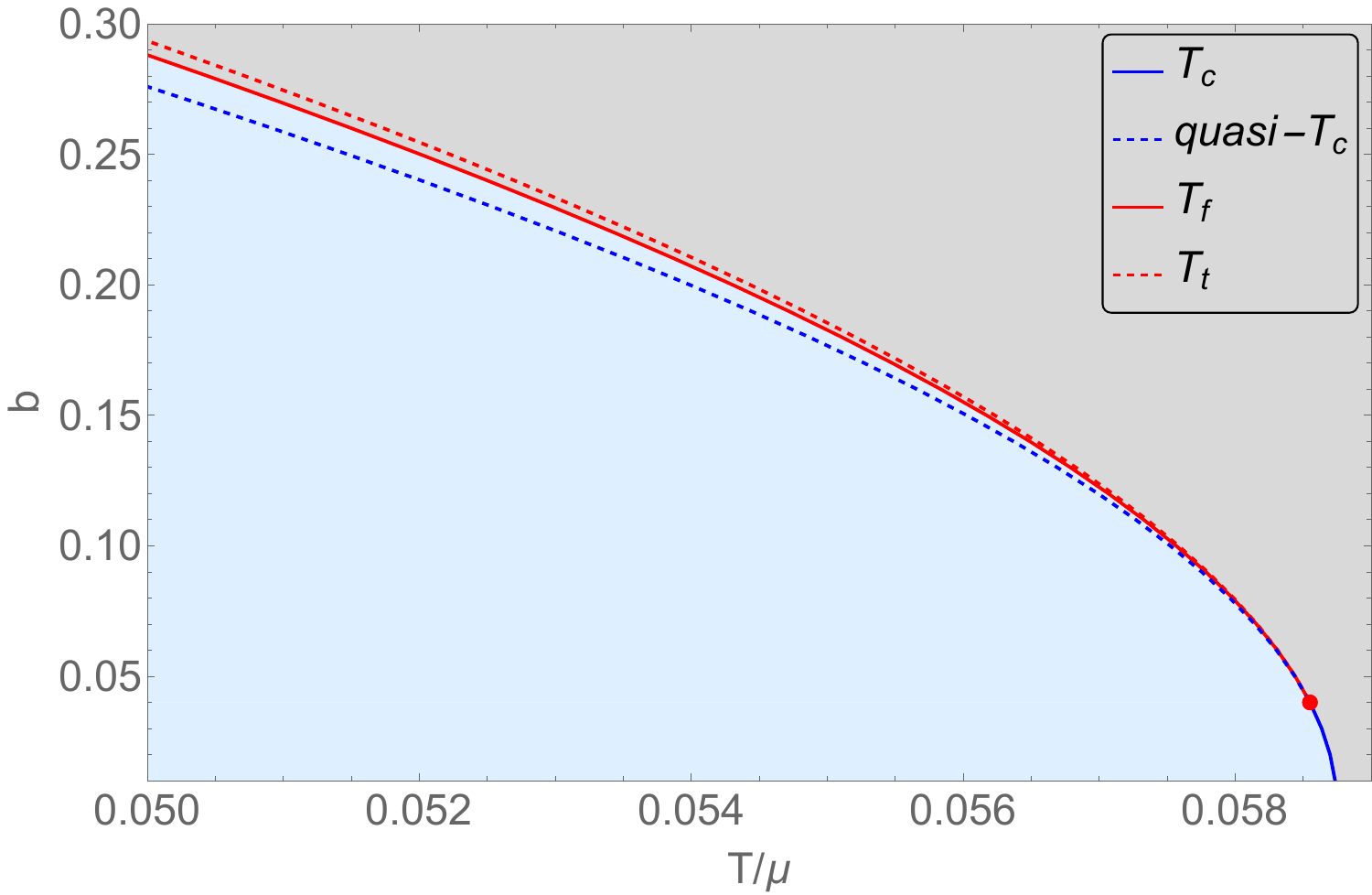}
		\caption{The phase diagram of temperature $T/\mu$ and back-reaction parameter $b$ with $\lambda=-0.75$ and $\tau=0.5$, the right panel is an enlarged view of a section of the left panel. The blue solid line represents the phase transition points of second-order phase transition. The blue dashed line represents the quasi-critical points of first-order phase transition at which the superfluid solution first appears. The red solid line represents the phase transition points of first-order phase transition. The red dashed line represents the turning points of first-order phase transition. The blue region corresponds to the superfluid phase.}\label{212phase}
	\end{figure}

	In the next step, we set $\lambda=-0.75$ and $\tau=0.5$, where $\lambda$ is slightly higher than the special value $\lambda_s$. In this case, the phase transition in the probe limit remains second-order. We present the condensates as well as the grand potential curves with increasing values of the back-reaction strength $b$ in Fig.~\ref{lambda-0.75b0.5}.
	We see that along with the increasing value of the back-reaction strength $b$ from the probe limit, the superfluid phase transition changes from the second-order to first-order. Interestingly, with the back-reaction strength continuously growing to a large enough value, the superfluid phase transition turns back to second-order. We further present a $b-T$ phase diagram in Fig.~\ref{212phase} to show this interesting behavior more clearly. From this phase diagram, we can see two red points at the ends of the first-order phase transitions, which indicate a non-monotonic control on the condensate near the critical point from the back-reaction strength similar to the fourth-power term with coefficient $\lambda$.
	
	\subsection{COW phase transitions and the supercritical superfluid}\label{chapter3.4}
	Following the above analysis of back-reaction effects on zeroth-order, first-order, and second-order phase transitions, we continue our study on the final case: the COW phase transitions. With $\lambda_s<\lambda<0$ and a small positive value of $\tau$, the condensate curve shows a COW configuration. At this time, the system undergoes a second-order superfluid phase transition from the normal phase, as well as a first-order phase transition between two superfluid phases with different values of condensate. It is confirmed that the first-order phase transition between superfluid phases in the COW case ultimately terminates at a critical endpoint when $\tau$ increases~\cite{Zhao:2024jhs}. Since the back-reaction strength brings in effects similar to the sixth-power term coefficient $\tau$, we set $\lambda=-0.2$ and $\tau=0.0072$ and present the condensate curves with increasing values of the back-reaction strength $b$ to see whether we are able to find the critical point.
	\begin{figure}[t]
		\center
		\includegraphics[width=0.46\columnwidth]{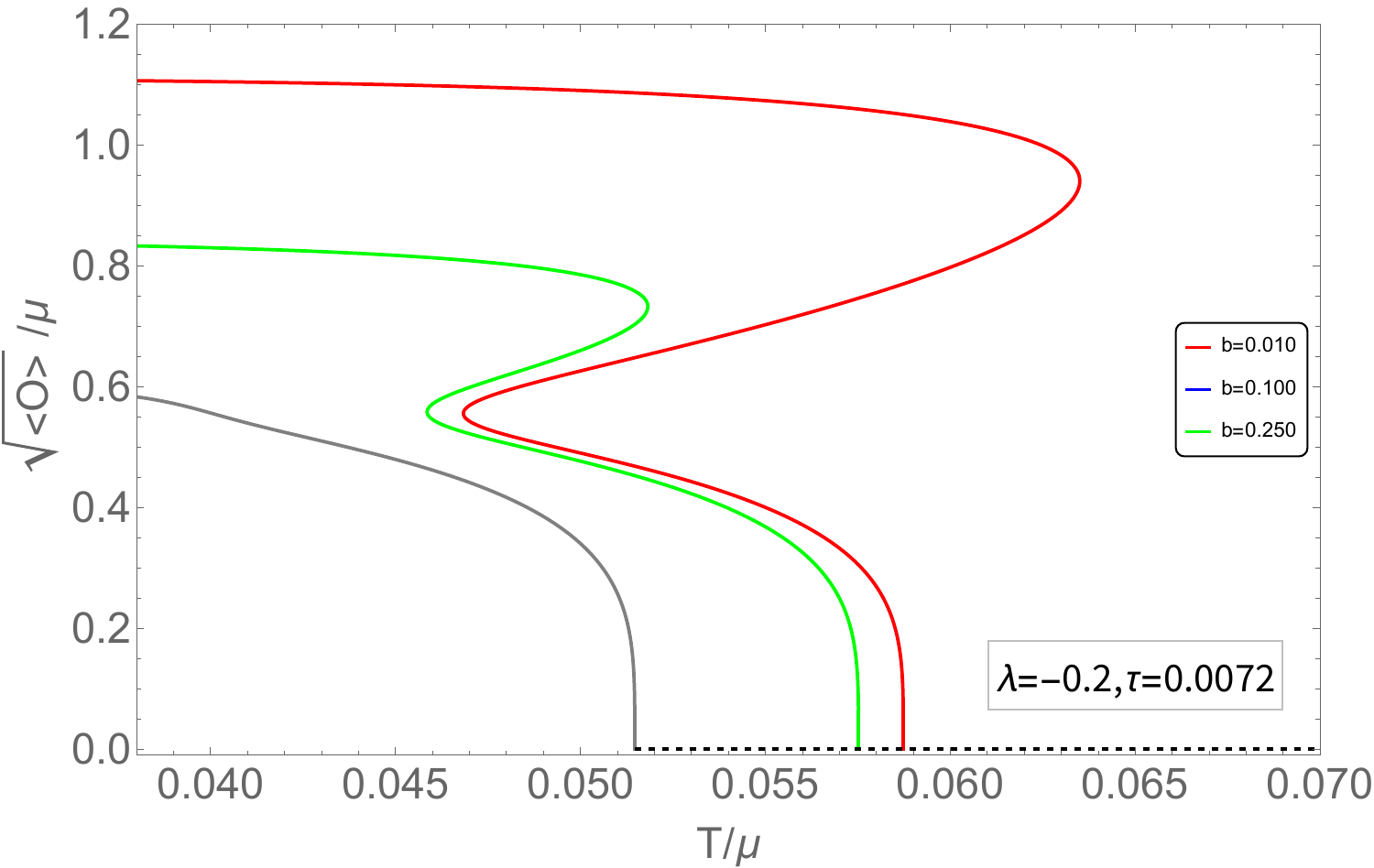}
		\includegraphics[width=0.49\columnwidth]{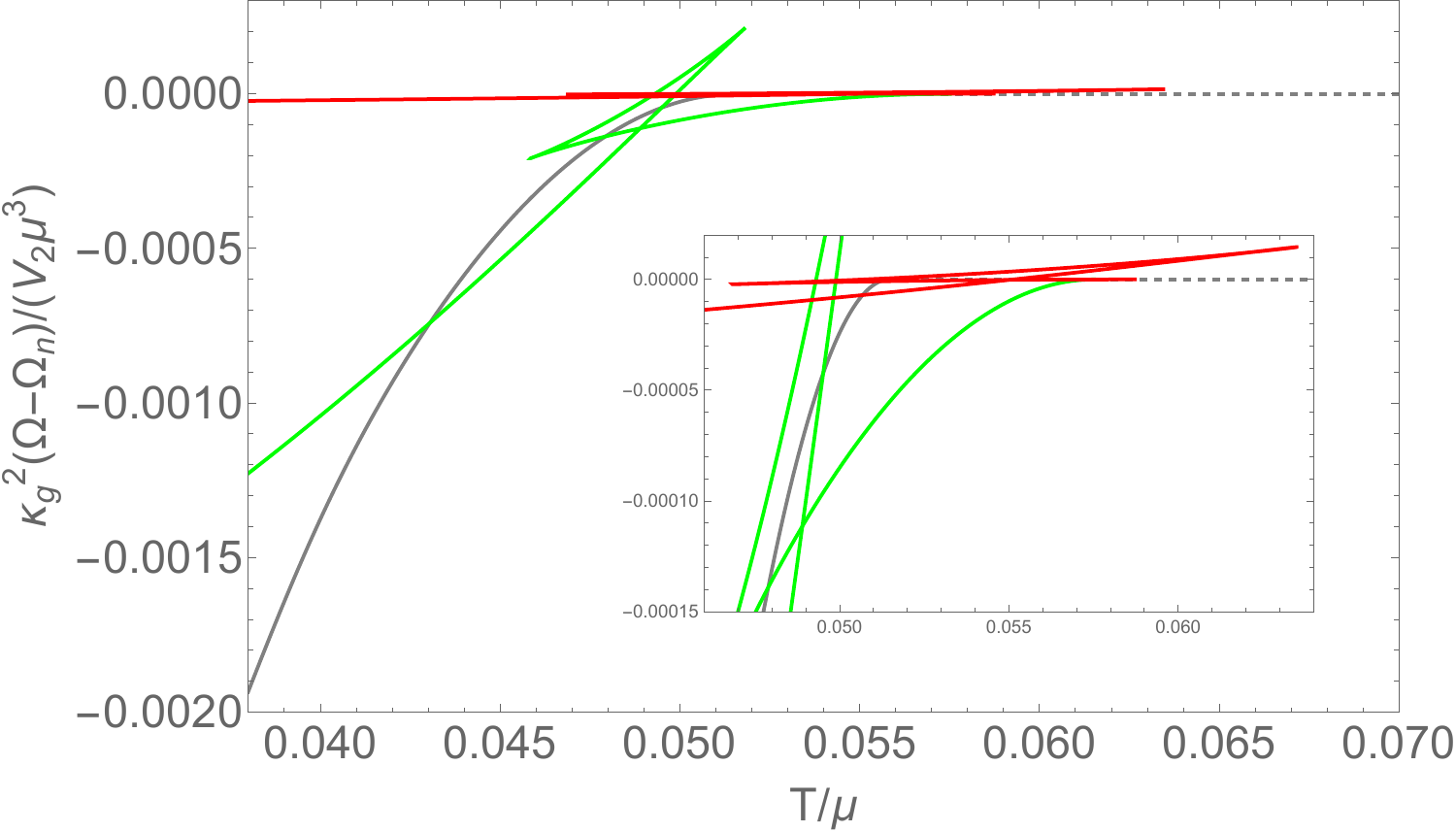}
		\caption{The dependence of the condensates as well as grand potential curves on the back-reaction strength $b$ with $\lambda=-0.2$ and $\tau=0.0072$. The left panel depicts the condensates, while the right panel represents the corresponding grand potential curves. The red, green, and gray lines represent the solutions with $b=0.010$, $0.100$, and $0.250$, respectively.}\label{lambda-02tau0o0072}
	\end{figure}
	
	We show the condensates as well as the grand potential curves with three values of the back-reaction strength $b$ in Fig.~\ref{lambda-02tau0o0072}. We can see from this figure that with increasing value of the back-reaction strength, the swallowtail region in the first-order phase transition becomes narrower and finally disappears while the condensate curve transitions to the typical second-order type.
	
\begin{figure}[t]
		\center
		\includegraphics[width=0.8\columnwidth]{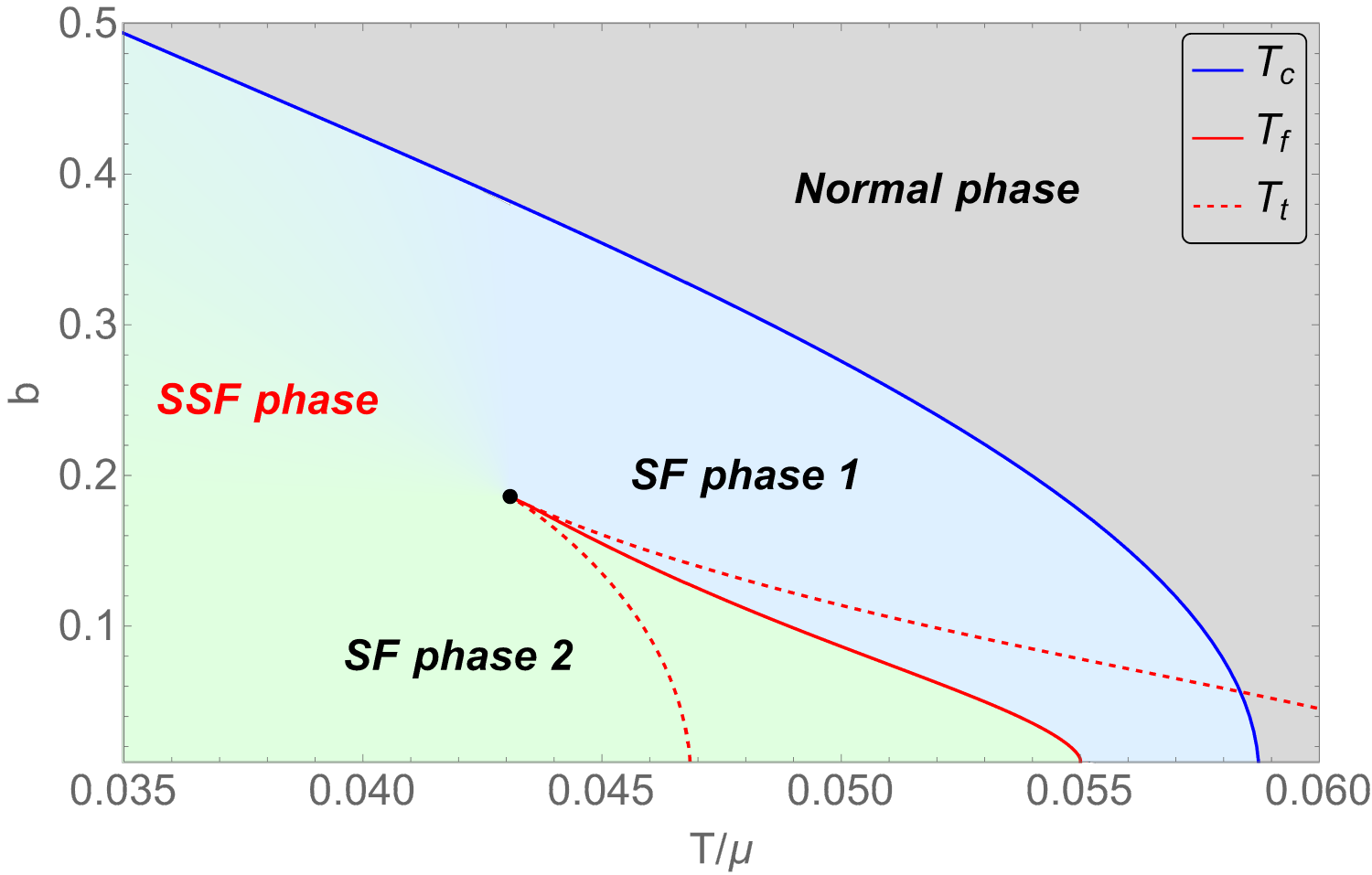}
		\caption{The $b-T$ phase diagram with $\lambda=-0.2$ and $\tau=0.0072$. The blue solid line represents the critical points of the second-order superfluid phase transitions.
			The red solid line represents the phase transition points of the first-order phase transitions between two different superfluid phases and the black point is the critical point at the end of these first-order phase transitions. Here SF represents superfluid and SSF represents supercritical superfluid. The red dashed lines are the turning points of the condensate curves in the first-order phase transitions.   }\label{superciritcalDiagram}
	\end{figure}

	Moreover, we construct the $b-T$ phase diagram from the phase transitions with various values of the back-reaction strength $b$ and present the results in Fig.~\ref{superciritcalDiagram}. In this phase diagram, we see a critical point at the end of the curve indicating the first-order phase transition points, beyond which is the supercritical region.
	%where the superfluid solutions with larger (SF phase 2) and smaller (SF phase 1) condensates are no longer distinguishable. 
	This phase diagram is similar to the one obtained by varying the value of $\tau$ instead of $b$, confirming our conjecture that the back-reaction strength induces effective coupling similar to the sixth-power term with coefficient $\tau$.

	\section{The influence of $\lambda$ and $\tau$ on phase transitions with finite back-reaction strength}\label{sec4}
	\subsection{Self-interaction terms}\label{b0}
	%我们从不考虑反作用的探子极限理论出发。
	
	\begin{figure}[t]
		\center
		\includegraphics[width=0.475\columnwidth]{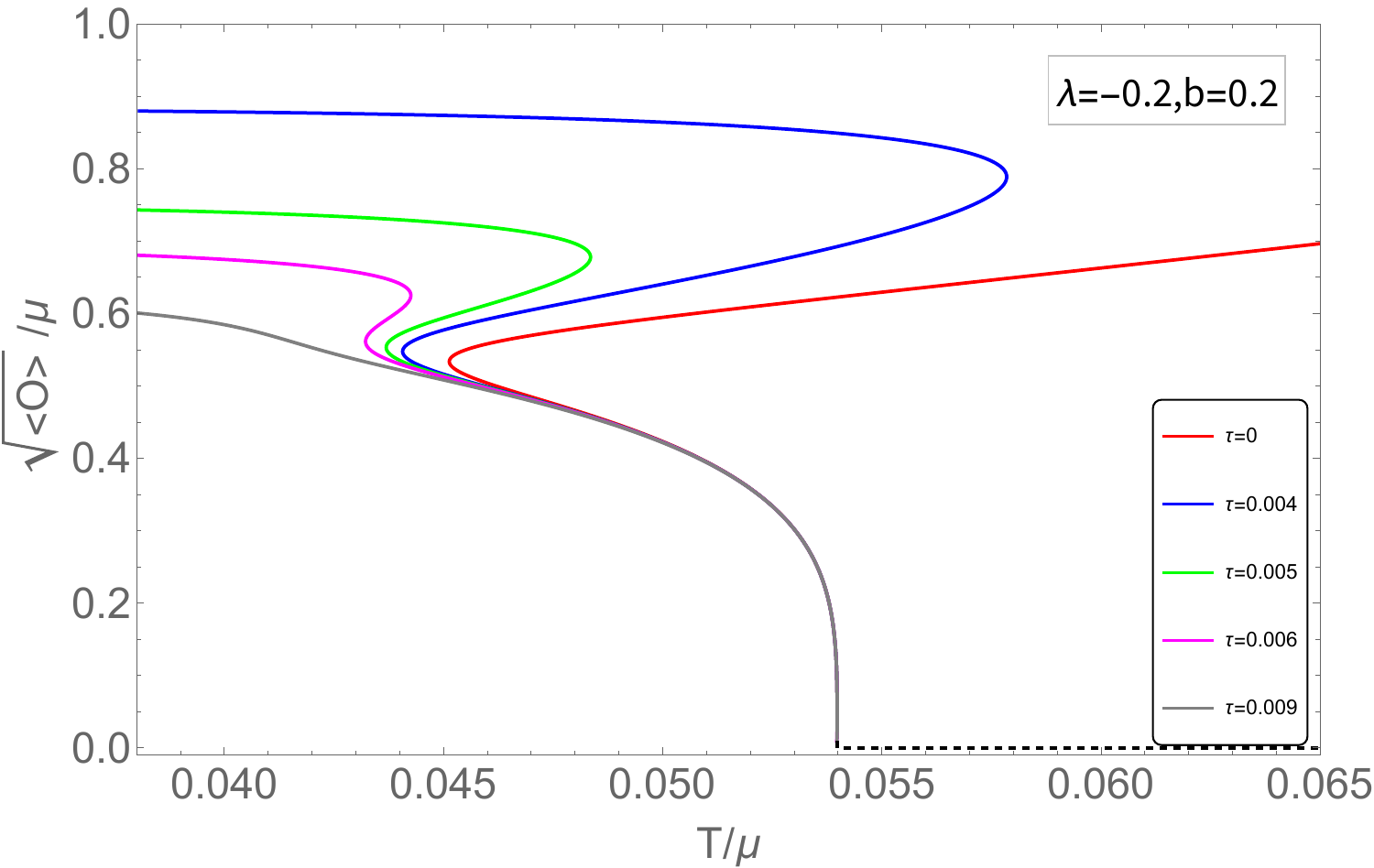}
		\includegraphics[width=0.49\columnwidth]{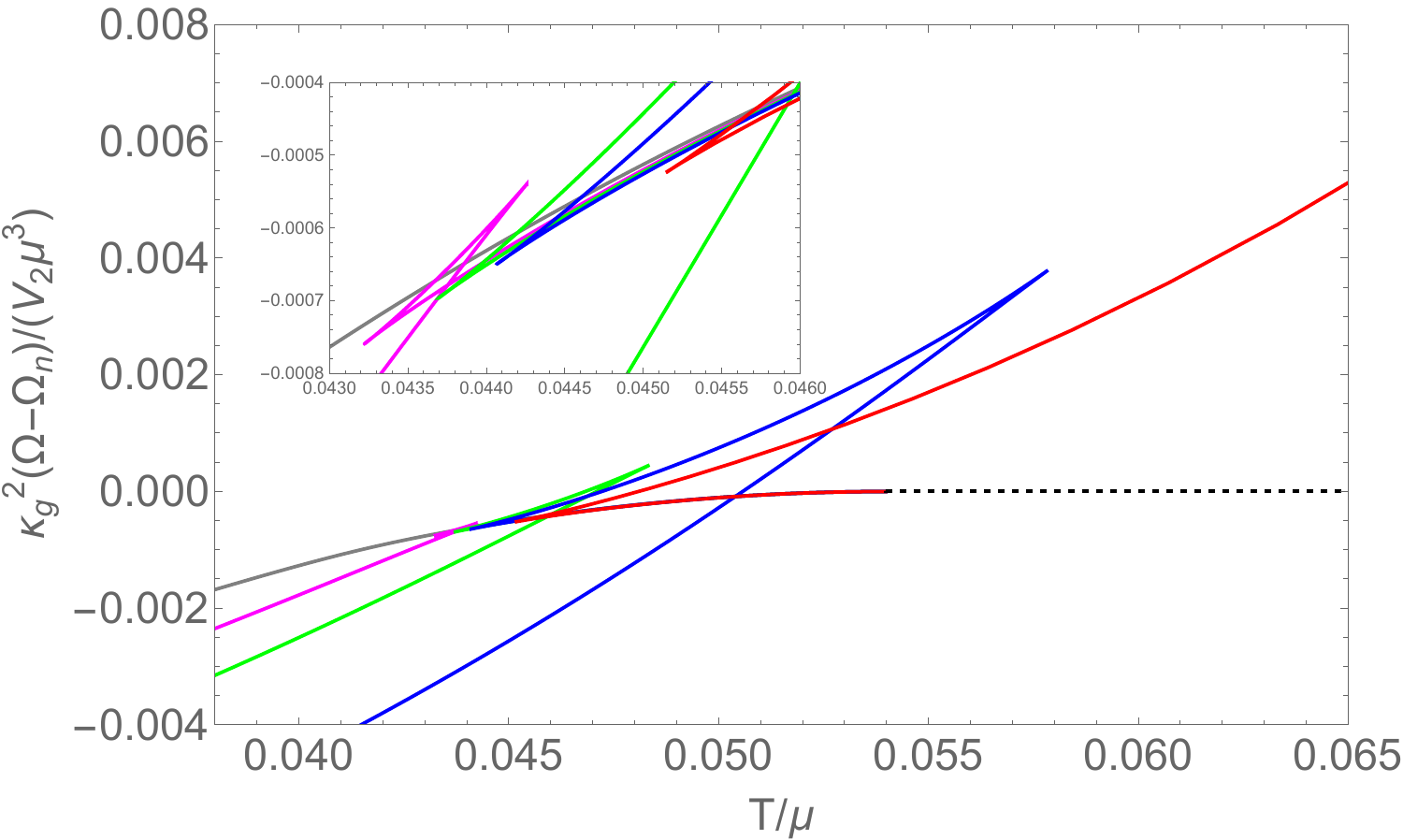}
		\caption{The dependence of the condensates as well as grand potential curves on the sixth-power coefficient $\tau$ with $\lambda=-0.2$ and $b=0.2$. The left panel depicts the condensates, while the right panel represents the corresponding grand potential curves. The red, blue, green, magenta, and gray lines represent the solutions with $\tau=0$, $0.004$, $0.005$, $0.006$, and $0.009$, respectively.
		}\label{lambda-02b0.2}
	\end{figure}
	%从图上来看，红色和蓝色的实线似乎都是零阶相变。但是理论上来讲，只要参数τ是一个正的并且非零的数，无论τ多么小，在psi足够大的时候τ总是会主导系统的稳定性。所以只有红色实线是零阶级相变，蓝色实线应该是一个一阶相变。为什么蓝色实线看上去是零阶相变的原因，是因为在τ足够小的时候，我们很难计算到足够大的psi使得系统再次转变回一阶相变，所以我们只能从理论分析的角度得出这个蓝色实线是一节相变这个结论。

	With the two non-linear term coefficients $\lambda$ and $\tau$, various phase transitions are realized in the probe limit~\cite{Zhao:2022jvs}, which presents powerful control of $\lambda$ and $\tau$ on the superfluid solutions. We have confirmed that this powerful control of $\lambda$ and $\tau$ still works in a similar way when the back-reaction strength is finite, and therefore we are able to tune the holographic superfluid phase transitions more accurately even with finite back-reaction strength.
	
	Fixing the finite back-reaction strength $b=0.2$, we illustrate the condensates as well as the grand potential curves with $\lambda=-0.2$ and different values of $\tau$ in Fig.~\ref{lambda-02b0.2}.
	From the two panels in Fig.~\ref{lambda-02b0.2}, we confirm that the effect of the increasing value of $\tau$ with finite back-reaction effect is the same as in the probe limit. With $\tau=0$, the phase transition is likely to be zeroth-order as shown by the red curves. However, as we have explained in Sect. \ref{subsection0th}, the red solid line is expected to turn back at very large condensate values to form the style of a first-order phase transition, due to the effective interaction introduced by the finite value of the back-reaction strength $b$. With the increasing value of $\tau$, the superfluid phase transitions change from first-order to second-order, leaving a critical point at the end of the line of first-order phase transition points, which is the same as in the probe limit. The region of the supercritical superfluid is also available in this situation beyond the critical point.
	
	%随着tau的值的增加，一阶相变的燕尾曲线会逐渐减小，最终在某个临界点消失。这种存在一阶相变并且在某个临界点消失的现象我们在第3.4章和（）中已经讨论过了，它的动力学和热力学特性非常具有普适性质。

\begin{figure}[t]
		\center
		\includegraphics[width=0.475\columnwidth]{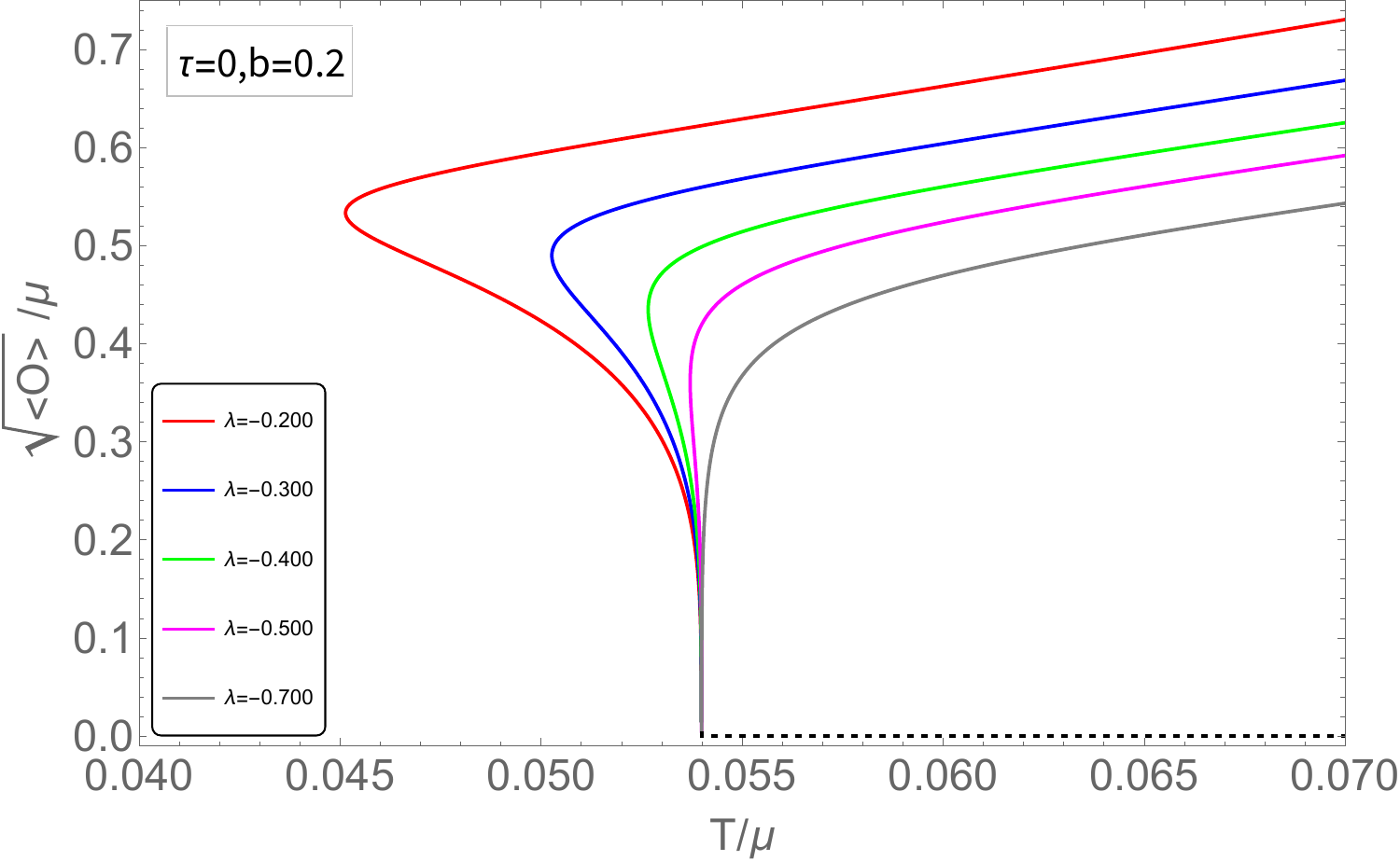}
		\includegraphics[width=0.49\columnwidth]{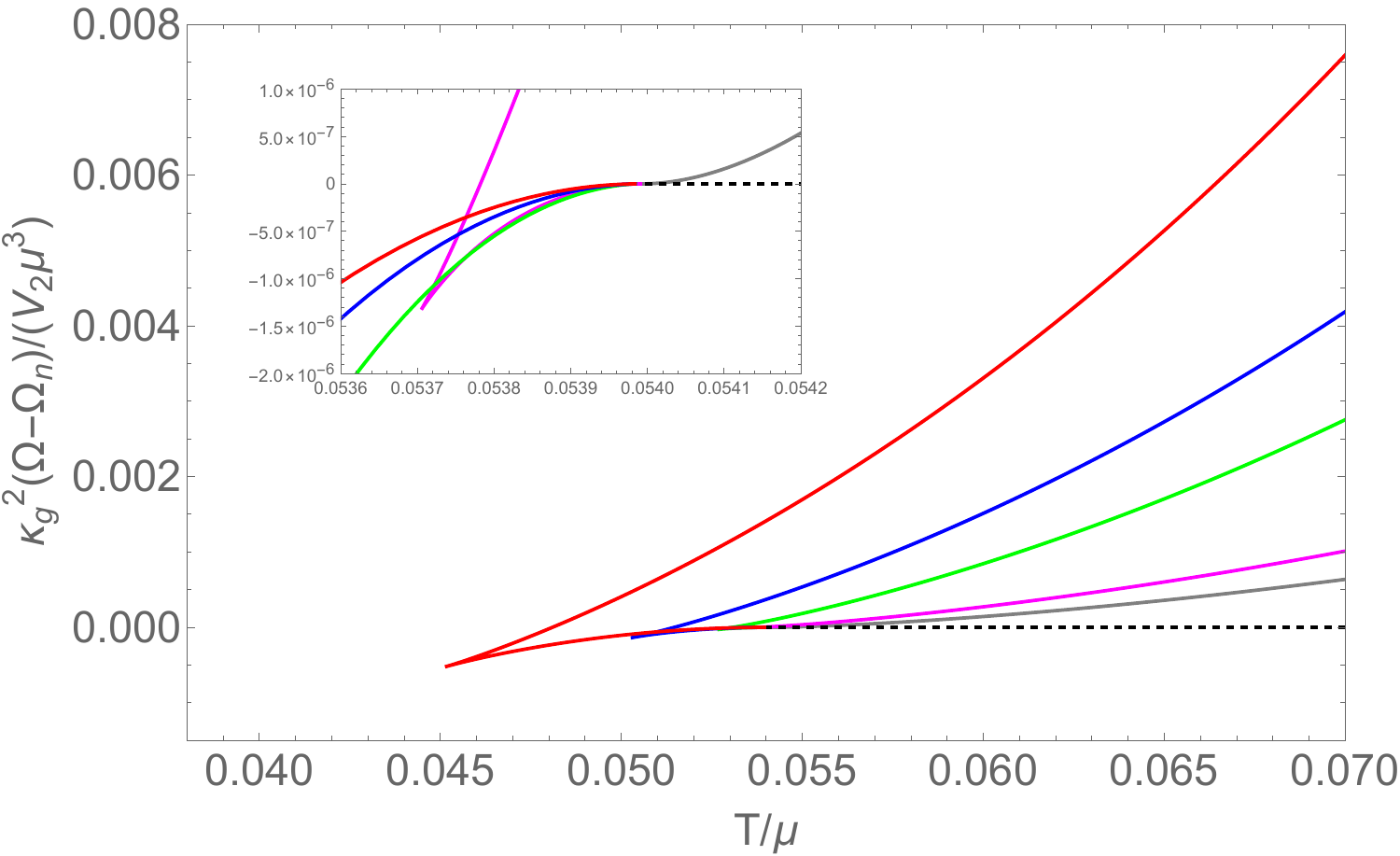}
		\caption{The dependence of the condensates as well as grand potential curves on the fourth-power coefficient $\lambda$ with $\tau=0$ and $b=0.2$. The left panel depicts the condensates, while the right panel represents the corresponding grand potential curves. The red, blue, green, magenta and, gray lines represent the solutions with $\lambda=-0.200$, $-0.300$, $-0.400$, $-0.500$, and $-0.700$, respectively.
		}\label{tau0b0.2}
	\end{figure}

	At finite back-reaction strength, the fourth-power term coefficient $\lambda$ also shows a similar influence on the superfluid phase transitions as in the probe limit. Compared to the sixth-power term coefficient $\tau$, $\lambda$ is more efficient in tuning the superfluid solutions near the critical point. 
	We show the condensates as well as the grand potential curves with increasing values of $\lambda$ in Fig.~\ref{tau0b0.2}, while fixing $\tau=0$ and a finite back-reaction strength $b=0.2$. We see clearly from the left panel that the red, blue, green, and magenta condensate curves grow leftward at the critical point, while the gray curve grows rightward. Notably, because a non-zero back-reaction strength introduces an effective coupling similar to a sixth-power term with positive value of $\tau$, all the condensate curves in Fig.~\ref{tau0b0.2} should eventually turn to grow leftward at sufficiently large condensate.
	
\begin{figure}[t]
		\center
		\includegraphics[width=0.8\columnwidth]{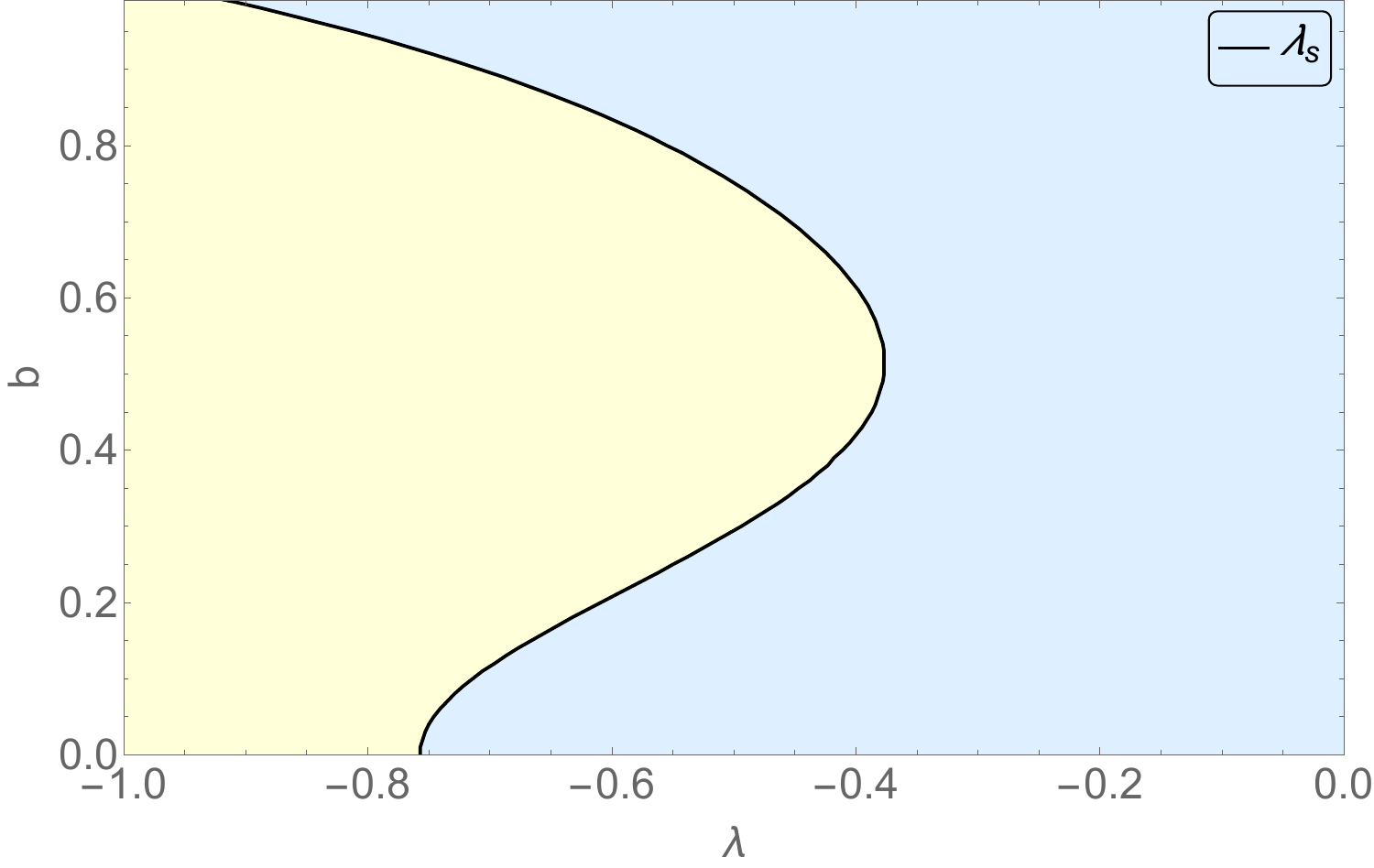}
		\caption{The dependence of the special value $\lambda_s$ on the back-reaction strength $b$. The blue region represents the parameter space where the condensates grow leftwards at the critical point, while the yellow region represents the parameter space where the condensates grow rightwards at the critical point.}\label{lambdaC_b}
	\end{figure}

	Similar to the case in the probe limit, we see a special value $\lambda_s$, which divides the different growth directions of the condensate curve, and is important for switching between the first-order and second-order superfluid phase transitions. This special value should depend on the value of the back reaction strength. And from the results in Sect.~\ref{1storderlambda}, the dependence of $\lambda_s$ on $b$ is non-monotonic. We plot the dependence of $\lambda_s$ on $b$ in Fig.~\ref{lambdaC_b}. In this plot, the black curve shows the dependence of the special value $\lambda_s$ on the back-reaction strength $b$, dividing this parameter space into the left yellow part dominated by the first-order superfluid phase transition and the right blue part dominated by the second-order superfluid phase transition.

	%我们在图9中给出了有限反作用下lambda对凝聚图的影响，其中红色，蓝色，绿色和品红色属于第一种情况，灰色属于第二种情况。值得注意的是，因为反作用不为零，所以理论上来讲，图八中所有的凝聚图都应该是一阶相变。只不过因为数值计算上的困难，我们没办法计算到足够大的psi。
	
	%但是，正如我们前面分析的那样，反作用会改变系统的相变类型，因此无法确定是否是零阶相变。所以这里我们将分为如下两种情况：第一种情况，凝聚值的凝聚温度从临界点出发一开始就是向着低于临界温度的方向生长（处于二阶相变或者是COW相变）。第二种情况，从一开始就向高于临界温度的方向生长（处于零阶相变或者是一阶相变）。
	
	\section{Conclusions and discussion}\label{secConclusions}
	In this work, we studied the phase transitions of the holographic s-wave superfluid model with the fourth- and sixth-power non-linear terms beyond the probe limit. This setup enables us to examine the effect of the increasing back-reaction strength on the various types of phase transitions, including the second-order, first-order, and zeroth-order, as well as the COW phase transitions.
	The second-order phase transition is the simplest, where only one phase transition from the normal phase to the superfluid phase occurs and the condensate curve grows as the temperature decreases. The zeroth-order phase transition
	in our holographic study is expected to take place when the condensate curve turns in the opposite direction with a finite condensate; below the temperature of this turning point, the superfluid solution no longer exists, and the system is compelled to revert to the normal solution. However, a previous study~\cite{Zhao:2022jvs} concluded that the zeroth-order phase transition, in general, is not able to occur. Compared to the second-order phase transition, the phase transition from the normal phase to the superfluid phase becomes first-order when the condensate curve first grows with increasing temperature at the critical point and then turns back with decreasing temperature. At this time, the free energy curve forms a swallowtail shape as illustrated in Fig.~\ref{lambda-0.78b0.4}. Finally, the COW phase transition mainly describes the double turning of the condensate curve, which is in principle a first-order phase transition between the two sections of superfluid phase transitions with larger and smaller condensate values, respectively, as illustrated in Fig.~\ref{lambda-02tau0o0072}. We are also able to confirm the universal control of the fourth- and sixth-power term coefficients $\lambda$ and $\tau$ on the phase transitions beyond the probe limit. We further give the dependence on the back-reaction strength $b$ of the special value $\lambda_s$, beyond which the condensate curve grows in an opposite direction at the critical point.
	
	In such a setup, we are also able to understand the effect of the finite back-reaction strength from an effective coupling point of view. It is already clear that increasing the back-reaction strength $b$ will deform the background metric, therefore, the critical temperature of the superfluid phase decreases. Following the condensates with increasing values of the back-reaction strength from the different phase transitions in the probe limit, we see that the back-reaction strength contains effective couplings similar to both the fourth-power and the sixth-power terms. This could be understood as meaning that with finite back-reaction strength, the condensate of the scalar field deforms the metric tensor, while the deformation of the metric tensor in turn affects the scalar field, which possibly brings in effective couplings similar to the fourth- and sixth-power terms. The effective coupling of the sixth-power term from the back-reaction strength seems positive, while the effective coupling of the fourth-power term might change sign with the increasing value of the back-reaction strength, as indicated by the non-monotonic dependence of $\lambda_s$ on $b$ in Fig.~\ref{lambdaC_b}.
	
	In Ref.~\cite{Zhao:2022jvs}, it is argued from the landscape analysis that as long as the sixth-power coefficient $\tau$ gets a positive value, no matter how small it is, this term will dominate in the region with a sufficiently large condensate and stabilize the system by bounding the thermodynamic potential landscape from below. If the back-reaction strength also brings in an effective sixth-power potential with a positive coefficient, we can also confirm the stability of the system from the landscape point of view, and expect a stable superfluid phase with a very large condensate at small values of the back-reaction strength $b$.
	
	There are many interesting topics to be further investigated in future studies. It is interesting to confirm the effective couplings brought by the back-reaction strength by expanding the Lagrangian to the linear level in the back-reaction strength $b$ to obtain the effective action on the background metric. Then it is straightforward to see how the back-reaction affects the holographic superfluid phase transitions analytically. With a finite value of the back-reaction strength, it is possible to consider the entanglement entropy, the complexity, as well as the black hole interior with various holographic superfluid phase transitions. From another perspective, crossovers in the supercritical region are frequently mentioned in recent studies (see,  \textit{e.g.}, Refs. \cite{DasBairagya:2019nyv,Sahay:2017hlq,Wei:2019yvs,Wei:2023mxw,Zhao:2025ecg,Xu:2025jrk,Wang:2025ctk}).
	This should be investigated in more detail in future studies. It is interesting to study the supercritical region in our $b-T$ phase diagram in Fig.~\ref{superciritcalDiagram} to promote the study on the holographic aspect of the gravity system.
	
	\vspace{\baselineskip} 
	\noindent\textbf{Acknowledgements}
	%ZYN would like to thank ... for useful discussions. He would also like to thank the organizers of ``...'' for their hospitality. 
	This work is partially supported by NSFC with Grant nos.12473001, 11965013, 11975072, 11881240248, 11565017, 11875102, and 11835009. ZYN is partially supported by Yunnan High-level Talent Training Support Plan Young $\&$ Elite Talents Project (Grant no. YNWR-QNBJ-2018-181). This work was supported by the National SKA Program of China (Grant nos. 2022SKA0110200
	and 2022SKA0110203), and the National 111 Project (Grant No. B16009).
	
	\vspace{\baselineskip} 
	\noindent\textbf{Data Availability Statement}
	This manuscript has no associated data. [Authors’ comment: Data sharing not applicable to this article as no datasets were generated or analysed during the current study.]
	
	\vspace{\baselineskip} 
	\noindent\textbf{Code Availability Statement}
	This manuscript has no associated code/software. [Author’s comment: Code/Software sharing not applicable to this article as no code/software was generated or analysed during the current study.]

	\bibliographystyle{apsrev4-1}
	\bibliography{holoptV2}
	
\end{document}